\documentclass[final]{siamonline250211}

\usepackage{amsmath,amssymb}
\usepackage{mathtools}
\usepackage{bbm}
\usepackage{graphicx}
\usepackage{xcolor}
\usepackage{float}
\usepackage{enumerate,enumitem}
\usepackage{algorithm,algpseudocode}
\algrenewcommand\algorithmicrequire{\textbf{Input:}}
\algrenewcommand\algorithmicensure{\textbf{Output:}}
\usepackage{booktabs}
\usepackage{cite}

\newcommand{\ind}{\mathbbm{1}}
\newcommand{\E}{\mathbb{E}}
\newcommand{\R}{\mathbb{R}}
\newcommand{\PP}{\mathbb{P}}
\newcommand{\Var}{\operatorname{Var}}

\newcommand{\Cov}{\operatorname{CoV}}
\newcommand{\cov}{\operatorname{Cov}}

\newcommand{\CV}{\operatorname{CV}}
\newcommand{\Id}{\operatorname{Id}}

\newcommand{\ve}{\varepsilon}
\newcommand{\tr}{\operatorname{tr}}
\newcommand{\bs}[1]{\boldsymbol{#1}}

\newcommand{\pecan}{pecan}
\newcommand{\Pecan}{Pecan}
\newcommand{\sundae}{pecanilla}
\newcommand{\Sundae}{Pecanilla}

\newcommand{\cv}[1]{\textcolor{blue!85!black}{#1}}

\newsiamthm{assumption}{Assumption}
\crefname{assumption}{Assumption}{Assumptions}
\newsiamremark{remark}{Remark}
\crefname{remark}{Remark}{Remarks}

\AtBeginDocument{
  \crefname{section}{\S\!\!}{\S\!\!}
  \Crefname{section}{Sec.\!}{Sec.\!}
  \crefname{subsection}{\S\!\!}{\S\S\!\!}
  \crefname{subsubsection}{\S\!\!}{\S\S\!\!}
  \Crefname{subsection}{Sec.\!}{Sec.\!}
}

\title{Beyond exponential tilting: sharp asymptotics and improved importance sampling for rare events\thanks{Version from \today.\funding{Partially supported by the U.S.\ Department of Energy, Office of Science Energy Earth Shot Initiative under Award number \#DE-SC0024721. 
}}}

\author{
Mingxin Li\thanks{Courant Institute School of Mathematics, Computing, and Data Science, New York University, New York, US (\email{ml6223@nyu.edu}, \email{timo.schorlepp@nyu.edu}, \email{stadler@cims.nyu.edu})}
\and Hans Reimann\thanks{Institut f{\"u}r Mathematik, Ruprecht-Karls-Universit{\"a}t Heidelberg, Heidelberg, Germany (\email{reimann@math.uni-heidelberg.de})}
\and Timo Schorlepp\footnotemark[2]
\and Georg Stadler\footnotemark[2]
}

\headers{Improved importance sampling for rare events}{Li, Reimann, Schorlepp, Stadler}

\begin{document}
\maketitle

\begin{abstract}
A classical approach for importance sampling (IS) of a rare event is to tilt the measure using its large deviation minimizer, thereby shifting the proposal mean towards the event set. While this tilt is log-efficient, we show that the coefficient of variation of the resulting estimator still grows like $\ve^{-1/4}$ in the small-noise limit $\ve\downarrow0$. Motivated by sharp large deviation asymptotics and the local geometry of the event set at the minimizer, we propose improvements using a control variate or an anisotropic, non-Gaussian proposal. We prove, up to truncations that hold with high probability, and confirm empirically in high-dimensional examples, that the former bounds the coefficient of variation and the latter makes it vanish, so that a bounded number of samples suffices. Our analysis addresses both the coefficient of variation of the different proposals and their necessary and sufficient sample size based on the Kullback--Leibler divergence of the IS target from the proposal; both yield the same rates in $\ve$. We demonstrate the resulting samplers, including a variant that blends the classical and the new proposal across noise levels, for a stochastic Korteweg--de Vries equation with thousands of random inputs.
\end{abstract}



\section{Introduction}
Direct Monte Carlo (MC) sampling of small probabilities is often prohibitively expensive. Yet, the estimation of small probabilities is needed for risk assessment in areas such as climate extremes and structural reliability. For rare events, most samples miss the event and add little information. Importance sampling (IS) mitigates this by sampling from a proposal distribution that concentrates on the rare-event region, followed by reweighting samples to keep the estimator unbiased. The success of IS depends crucially on how this proposal distribution is chosen; a poor choice can perform worse than direct MC, or even lead to an estimator with infinite variance~\cite{glasserman-wang:1997}.

Large deviation theory (LDT) provides a natural starting point for rare event importance samplers~\cite{siegmund:1976,vanden-eijnden-weare:2012,bucklew:2013,tong-vanden-eijnden-stadler:2021}. In the small-noise limit, it identifies the most likely realization of the rare event, the \emph{instanton}~\cite{grafke-grauer-schaefer:2015} (the \emph{design point} in reliability engineering~\cite{breitung:1984}). A standard LDT-based sampler shifts the mean toward the instanton---an exponential (``Cram{\'er}'') tilt of the measure. We call this vanilla IBIS (``instanton-based importance sampling'', after~\cite{ebener-margazoglou-friedrich-etal:2019}; ``vanilla'' as in ``plain''). Although this tilt renders the sampler log-efficient under standard assumptions~\cite{glasserman-wang:1997,bucklew:2013}, it ignores the curvature of the event boundary and the anisotropic fluctuations around the instanton, so its coefficient of variation still diverges in the small-noise limit. This issue has gained attention recently~\cite{en-zhang-huan:2023,katsevich-katsevich:2025}, and
the computational and theoretical tools to address it have only just become available, as has a Kullback--Leibler (KL) divergence-based theory of IS sample size
that we evaluate rigorously here (see \cref{sec:literature-review}).
The purpose of this paper is to understand this behavior and to develop and analyze improved, dimension-robust sampling methods that exploit the local shape of the event boundary. This requires a sharper analysis than the usual log-efficiency of IS~\cite{bucklew:2013}, focusing instead on sharp asymptotics for the coefficient of variation (CoV). We also discuss the KL divergence between the optimal IS distribution and our LDT-based methods, as suggested in \cite{chatterjee-diaconis:2018} as a proxy for the necessary and sufficient sample size.

\begin{figure}
    \centering
    \includegraphics{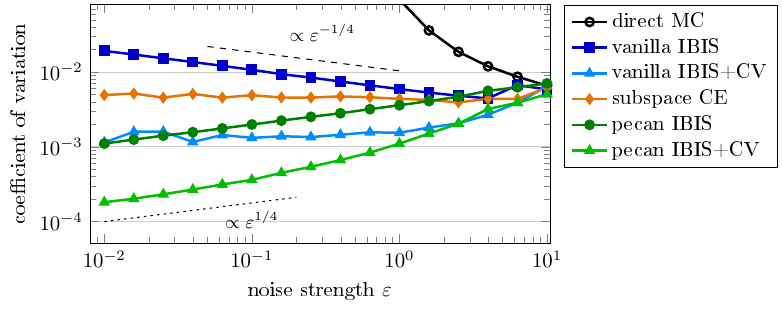}
    \vspace{-2ex}
    \caption{Coefficient of variation versus noise strength $\ve$ for six samplers applied to the SDE~\eqref{eq:2d-sde-intro}. Each estimate uses $10^5$ trajectories; subspace CE splits them evenly between training (8D subspace) and testing. As $\ve$ decreases, the event becomes rarer; estimated probabilities are $1.9 \times 10^{-1}$, $1.2 \times 10^{-3}$, $7.3 \times 10^{-7}$, $3.7 \times 10^{-22}$, and $5.0 \times 10^{-203}$ for $\ve = 10$, $1$, $10^{-0.4}$, $0.1$, and $0.01$, respectively.}
    \label{fig:cov-comparison}
\end{figure}

\subsection{Motivating example: coefficient of variation of six samplers}
\label{sec:intro-sde}
We consider the two-di\-men\-sional stochastic differential equation (SDE)~\cite{schorlepp-tong-grafke-etal:2023} for $(x_t,y_t)_{t\in[0,1]}$
\begin{equation}
\begin{cases}
\mathrm{d}x_t = (-x_t - x_t y_t)\,\mathrm{d}t + \sqrt{\ve} \,\mathrm{d}B_t^x\,,\\
\mathrm{d}y_t = (-4y_t + x_t^2)\,\mathrm{d}t + \frac{1}{2}\sqrt{\ve}\,\mathrm{d}B_t^y\,,
\end{cases}
\label{eq:2d-sde-intro}
\end{equation}
with $x_0=y_0=0$.
Here, $B^x$ and $B^y$ are independent one-dimensional Brownian motions, and $\ve>0$ sets the noise strength. We estimate the probability of the event $x_1+2y_1\ge 3$, which becomes increasingly rare as $\ve$ decreases since the origin is a stable fixed point of the deterministic dynamics. Discretizing~\eqref{eq:2d-sde-intro} with 2000 Euler--Maruyama steps yields 4000 independent Gaussian increments, turning the problem into a finite but high-dimensional one. For more on discretization and sampler implementation, see the more complex example in \cref{sec:kdv}.

\Cref{fig:cov-comparison} shows the CoV of six estimators across 16 noise strengths, $\ve \in [10^{-2}, 10^1]$. The results in \cref{fig:cov-comparison} can be reproduced using the scripts available in~\cite{pecan-ibis-sde}. The CoV, the standard deviation of an estimator divided by its mean, measures the relative sampling error. The six methods are as follows; the first three are existing approaches, the last three are proposed in this paper.

\begin{figure}
    \centering
    \includegraphics[width=0.85\linewidth]
    {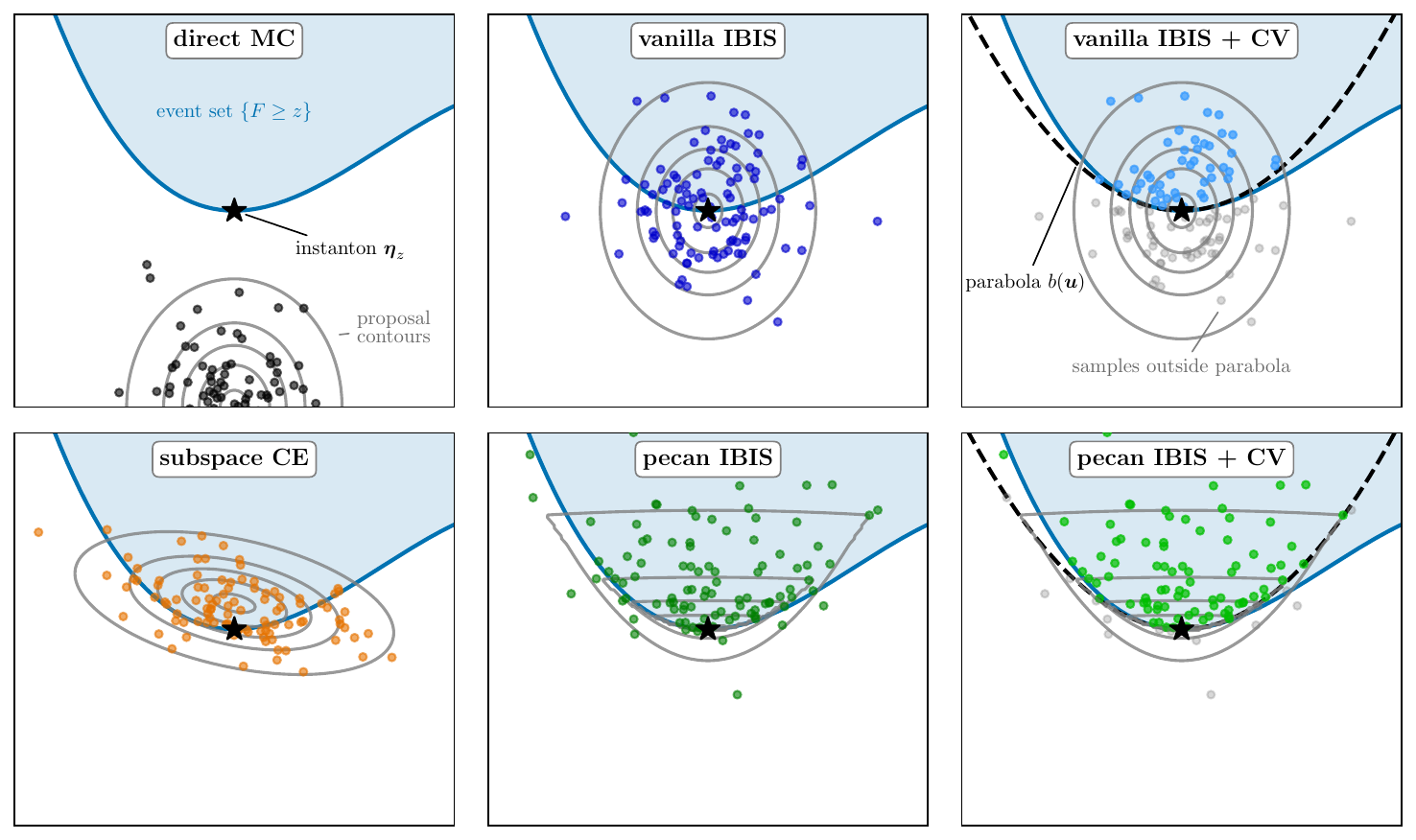}\vspace{-2ex}
    \caption{Sketch of IBIS family proposals in a two-dimensional toy problem. See \cref{sec:problem-setup} for a detailed explanation and formal definitions. 
    }
    \label{fig:sampler-sketch}
\end{figure}

\begin{enumerate}[leftmargin=1.5em]
\item \textbf{Direct MC} becomes unusable for small $\ve$ because almost no samples reach the event.
\item \textbf{Vanilla IBIS} shifts the mean of the distribution to the instanton~\cite{siegmund:1976}. While it greatly improves upon direct MC~\cite{bucklew:2013,ebener-margazoglou-friedrich-etal:2019,tong-vanden-eijnden-stadler:2021}, its CoV increases as $\ve$ decreases, so the sample size needed to keep error constant must grow.
\item \textbf{Subspace CE (cross-entropy)}\footnote{The plotted CoV in \cref{fig:cov-comparison} is the median over 20 independent repetitions because subspace CE occasionally produces large CoV outliers; the underlying cause for this is discussed in \cref{sec:conclusion} in a one-dimensional example.} is included for comparison. CE-based samplers~\cite{uribe-papaioannou-marzouk-etal:2021,tong-stadler:2023,beh-morio-simatos:2025} tune a (typically Gaussian) proposal to approximate the optimal IS distribution. We use the method from~\cite{tong-stadler:2023}, which modifies the standard IBIS proposal
within a low-dimensional, LDT-informed subspace while remaining Gaussian. In \cref{fig:cov-comparison}, it improves on vanilla IBIS and yields a constant CoV as $\ve \downarrow 0$. 
\item \textbf{Vanilla IBIS+CV (control variate)} keeps the vanilla IBIS estimator but corrects it with a control variate~\cite{glasserman:2004}, namely the analogous estimator for the quadratic approximation of the event boundary, whose mean is known in closed form~\cite{breitung:1984,tong-vanden-eijnden-stadler:2021,schorlepp-tong-grafke-etal:2023}. Since the two are strongly correlated for rare events, this greatly reduces the CoV.
\item \textbf{\Pecan\ IBIS} modifies the proposal by using
the theoretically known limiting exponential behavior along the normal direction pointing into the event set (referred to as ``above'' the parabola; cf.~\cref{fig:sampler-sketch}) and applying a curvature correction in the tangential directions; that is, a \textbf{P}iecewise-\textbf{E}xponential, \textbf{C}urvature-corrected \textbf{AN}isotropic proposal.
While a matter of taste for ice cream, \pecan\ does improve upon vanilla in our setting: its CoV decreases monotonically as $\ve\downarrow 0$.
\item \textbf{\Pecan\ IBIS+CV} combines the \pecan\ IBIS proposal with the quadratic control variate. It gives the smallest CoV in \cref{fig:cov-comparison}.
\end{enumerate}
In this paper, we provide a small-noise asymptotic analysis of methods 2, 4, and 5 in terms of their CoV, and methods 2 and 5 in terms of their KL divergence to the optimal IS distribution; a rigorous analysis of method 6 remains open.
The key takeaway from \cref{fig:cov-comparison} is that identifying the instanton alone does not yield a low-variance estimator at fixed sample size in the small-noise limit. While all of the above importance samplers improve upon standard MC, the local geometry of the event boundary and how samples concentrate near it also matter. The methods proposed in this paper leverage this local information to improve upon vanilla IBIS in the small-noise sense. All of the methods considered in this paper are sketched in \cref{fig:sampler-sketch}, and we define them in more detail in \cref{sec:problem-setup}.

\subsection{Main questions and contributions}
The first question is whether the growing CoV of vanilla IBIS -- which we show to be $\sim \text{const} \times \ve^{-1/4}$, in agreement with \cref{fig:cov-comparison} -- can be reduced. 
We show that a control variate based on the quadratic approximation of the event boundary removes the leading growth in the variance, so that the CoV of the resulting estimator remains bounded in the LDT limit $\ve \downarrow 0$ (for a truncation of the estimator that is inactive with high probability).

The second question is whether the proposal distribution itself can be improved. \Pecan\ IBIS uses the limiting exponential distribution in the direction normal to the event boundary and corrects for curvature in the tangential directions. It adds a sampling branch below the local quadratic boundary to cover 
the mismatch with the true event set. We derive the leading variance and leading-order tuning 
for this lower branch, resulting in CoV $\sim \text{const} \times \ve^{1/4}$ as $\ve \downarrow 0$ for the estimator localized to a fixed tangential neighborhood of the instanton.
Our KL analysis requires no such localization.
We discuss implementation details and large-$\ve$ performance of \pecan\ IBIS in \cref{sec:kdv}.

Finally, we combine \pecan\ IBIS with control variates. The resulting estimator attains the lowest CoV in our two-dimensional SDE example. Our current theoretical results analyze the small-noise limit of methods 2, 4, and 5 separately. 

Our analysis of sampler asymptotics uses two approaches: the CoV and a recent framework that treats the KL divergence between the proposal and the optimal biasing distribution as the primary driver of sample size~\cite{chatterjee-diaconis:2018}. In the small noise limit $\ve \downarrow 0$, both yield consistent sample-size rates, differing only in constants and their necessary assumptions. Our adaptation of~\cite{chatterjee-diaconis:2018} to the LDT regime may be of independent interest.

\subsection{Related literature}
\label{sec:literature-review}
Three developments make our analysis timely. First, instantons and LDT prefactors can now be computed at the scale of discretized stochastic partial differential equations (SPDEs)~\cite{tong-vanden-eijnden-stadler:2021,schorlepp-tong-grafke-etal:2023,schorlepp-grafke:2025}. Thus, the ingredients our methods are based on---gradients and dominant Hessian eigenpairs at the instanton---are available in practice. Second, the inadequacy of the plain exponential tilt has been noted independently~\cite{maes-breitung-dupuis:1993,en-zhang-huan:2023,katsevich-katsevich:2025} (see below). Finally, IS sample-complexity theory based on statistical divergences (rather than moments)~\cite{chatterjee-diaconis:2018,sanz-alonso:2018,
aeckerle-willems-klebanov-weissman:2026} and intrinsic dimensions \cite{agapiou-papaspiliopoulos-sanz-alonso-etal:2017,beh-shadmi-simatos:2025} has emerged in the last decade. Small-noise rare-event sampling, where every quantity has an explicit expansion, is a natural application for this theory.
Note that we analyze the small-noise limit $\ve \downarrow 0$ because it yields explicit speeds and a cleaner setting. Alternatively, one could study the extreme-event limit $z \to \infty$ for a fixed measure~\cite{dematteis-grafke-vanden-eijnden:2019,tong-vanden-eijnden-stadler:2021}.

Our numerical experiments consider high-dimensional discretizations of (small-noise) SDEs and well-posed SPDEs; Freidlin--Wentzell theory~\cite{freidlin-wentzell:2012} provides the large deviation principle for SDEs and semi-discretized SPDEs. Throughout, we restrict ourselves to ``large excursion events'', as opposed to metastable transitions~\cite{bouchet-reygner:2022,heller-limmer:2024}, where a continuous family of instantons exists due to time translation invariance.
Although sharp finite-dimensional asymptotics (our theoretical setting) are well understood~\cite{bleistein-handelsman:1975,iltis:1995}, the standard IBIS approach has not, to our knowledge, been analyzed with sharp asymptotics or improved systematically, as we do here.

Prior work has sought to improve standard IBIS. In \cite{vanden-eijnden-weare:2012}, the authors propose an asymptotically optimal sampler for small-noise SDEs with state-dependent (closed-loop) drift, but it requires re-solving the instanton problem along the trajectory. Our methods use a single precomputed instanton and are easy to implement when forward-map gradients are available, as is typical when the instanton is found via gradient-based optimization. 
The paper~\cite{katsevich-katsevich:2025} approximates the conditional law of the rare event by an exponential distribution above the quadratic boundary approximation and suggests it as an IS proposal, but without analysis or numerics. Since it assigns no mass below the parabola, where the event may extend, the resulting estimator would be biased. Our \pecan\ IBIS adds the missing bottom branch.

The techniques we build on are closely related to reliability engineering, where finite-dimensional LDT corresponds to the Second-Order Reliability Method (SORM)~\cite{breitung:1984,derkiureghian-stefano:1991,rackwitz:2001,hu-mansour-olsson-etal:2021}, and the instanton is called the ``design point.'' 
Although SORM-based control variates were recently proposed in~\cite{en-zhang-huan:2023}, they lack the rigorous asymptotic analysis we provide here. Similarly, \pecan\ IBIS corresponds to~\cite{maes-breitung-dupuis:1993} and is related to line sampling~\cite{schueller-iwo-pradlwarter:2004} and radial IS~\cite{melchers:1990}.

Our methods and guarantees require gradients, a sufficiently smooth event boundary~\cite{iltis:1995}, and a unique dominating instanton~\cite{bucklew:2013}. This global assumption is hard to verify in high dimensions; if the point is not truly dominating, IBIS can even increase variance versus standard MC~\cite{glasserman-wang:1997}. When these geometric assumptions do not hold or gradients are unavailable, alternatives include sequential MC/subset simulation~\cite{au-beck:2001,cerou-del-moral-furon-etal:2012,straub-papaioannou:2015,wagner-latz-papaioannou-etal:2020}, 
Markov chain MC~\cite{eshra-papakonstantinou:2025}, variational methods~\cite{friedli-ginsbourger-doucet-etal:2025}, and ensemble methods~\cite{wagner-papaioannou-ullmann:2022,ehre-papaioannou-straub:2023,chakraborty-harris-klein-etal:2025}. These gradient-free methods are robust but typically lack the strong efficiency guarantees of LDT-based samplers.

\subsection{Organization}
\label{sec:organization}
In \cref{sec:problem-setup} we formulate the rare-event sampling problem, review sharp LDT asymptotics in \cref{sec:ldt-setup}, and define our importance-sampling family in \cref{sec:ibis-family}. \Cref{sec:theory-cov} analyzes their small-noise CoV and explains \cref{fig:cov-comparison}, while \cref{sec:theory-kl} evaluates sampler quality via KL divergence to the optimal distribution. \Cref{sec:kdv} presents a numerical study of the stochastic Korteweg--de Vries equation
demonstrating our method's scalability. We conclude in \cref{sec:conclusion}; most proofs are in Appendices \ref{app:proofs-sec3} and \ref{app:proofs-sec4}.

\section{Problem formulation and the family of IBIS estimators}
\label{sec:problem-setup}
This section presents the large-deviation analysis of small-noise rare events and the needed assumptions. In \cref{sec:ibis-family}, we introduce instanton-based IS methods, whose asymptotic properties are analyzed in \cref{sec:theory-cov,sec:theory-kl}.

\subsection{Small-noise rare-event problem}
\label{sec:ldt-setup}
In this section, we formalize the rare event problem of interest and state the associated assumptions. Our notation and setup follow~\cite{schorlepp-tong-grafke-etal:2023}. Let $\bs \xi$ be a standard Gaussian random vector in~$\R^N$ with $N < \infty$ fixed\footnote{This may arise, e.g., from discretization of a stochastic process or random field; see \cref{sec:conclusion} for a brief discussion of an extension to infinite dimensions.
}. 
Let $F: \R^N \to \R$ be a map from the random parameter vector to a scalar quantity of interest. For $\ve > 0$, define the small-noise random input $\bs \eta_\ve = \sqrt{\ve}\bs \xi$. It follows that $\bs \eta_\ve \sim {\cal N}(\bs 0, \ve \operatorname{Id}_N)$. For a fixed threshold~$z$ with $z > F(\bs 0)$,
we are interested in the rare event probability
\begin{align}
P_\ve(z) := \PP\left(F \left(\bs\eta_\ve \right) \geq z\right) = (2\pi \ve)^{-N/2} \int_{F(\bs\eta) \geq z} \exp\left( - \frac{\Vert \bs\eta \Vert^2}{2\ve} \right)\,\mathrm{d}\bs\eta.
\label{eq:tail-prob-def}
\end{align}
Since $\bs\eta_\ve$ converges to zero in probability as $\ve \downarrow 0$, the probability $P_\ve(z)$ also approaches zero, and the event becomes increasingly rare in the small-noise limit. 
We use LDT to describe $P_\ve(z)$ as $\ve \downarrow 0$. The central idea is that, among all realizations in the event set, the one with the smallest ``cost'' gives the leading contribution to the rare-event probability. 
We define the large-deviation rate function associated with the event as\footnote{Note that the mapping $z \mapsto I(z)$ that this defines indeed coincides with the large-deviation rate function
in the usual sense~\cite{dembo-zeitouni:1998}, of the family of one-dimensional random variables $(F(\bs \eta_\ve))_{\ve > 0}$ for any $z > F(\bs 0)$. This motivates our notation and terminology, despite the fact that we consider $z$ fixed throughout this paper.} 
\begin{equation}
    I(z) := \inf_{\bs\eta: F(\bs\eta) \geq z} \frac{1}{2}\Vert \bs\eta \Vert^2.
    \label{eq:rate-func-def}
\end{equation}
Throughout the paper, we impose the following necessary and standard assumptions \cite{iltis:1995, tong-vanden-eijnden-stadler:2021,schorlepp-tong-grafke-etal:2023} to ensure the validity of the leading-order sharp LDT asymptotics for $P_\ve(z)$.
\begin{assumption}[for sharp LDT]
\label[assumption]{assump:standard-ldt}
We assume that
\begin{enumerate}[label=(\roman*), leftmargin=*]
    \item the parameter-to-event map $F$ is three times continuously differentiable on $\R^N$, with locally Lipschitz third derivative.
    \item there exists a unique global minimizer,
    which we denote by $\bs \eta_z$ and call ``instanton'', for the optimization problem~\eqref{eq:rate-func-def}.
    \item the instanton $\bs \eta_z$, by (i) and (ii), lies on the event boundary, $F(\bs \eta_z) = z$, and we further assume the non-degeneracy condition $\nabla F(\bs\eta_z) \neq 0$.
    \item the standard second-order sufficient optimality condition for~\eqref{eq:rate-func-def} holds at $\bs \eta_z$.
\end{enumerate}
\end{assumption}
By definition, we then have $I(z) = \frac{1}{2}\Vert \bs\eta_z \Vert^2$, and
under the large deviation principle~\cite{dembo-zeitouni:1998,bucklew:2013},
\begin{equation*}
    -\lim_{\ve \downarrow 0} \ve \log P_\ve(z) = I(z) \iff P_\ve(z) = \exp\left( -I(z) / \ve+ o(\ve^{-1})\right).
\end{equation*}
The first order optimality condition for~\eqref{eq:rate-func-def} gives
\begin{equation}
\bs\eta_z = \lambda_z \nabla F(\bs\eta_z)
\label{eq:first-order-opt}
\end{equation}
for a Lagrange multiplier $\lambda_z > 0$. Denote the distance from the origin to the instanton by $r := \Vert \bs\eta_z \Vert = \sqrt{2I(z)}$, and denote the unit normal direction by $\bs e_z := \bs\eta_z / r$. The orthogonal projection onto the tangent space $\bs\eta_z^\perp$ of the event boundary is $\text{pr}_{\bs\eta_z^\perp} := \operatorname{Id}_N - \bs e_z \otimes \bs e_z$. The projected Hessian $H_z := \text{pr}_{\bs \eta_z^\perp} \nabla^2 F(\bs \eta_z) \text{pr}_{\bs \eta_z^\perp}$ describes the local curvature of the event boundary near the instanton. We abbreviate
$G_z := \operatorname{Id}_{\bs \eta_z^\perp} - \lambda_z H_z$, which is used in the probability asymptotics and in the curvature transformation for \pecan\ IBIS. Under \cref{assump:standard-ldt} (iv), the matrix $G_z$ is positive definite.
Lastly, for scalar $\mu$, we define
\begin{equation}
C_{\rm SORM}(z, \mu) := \left[ 2I(z) \det \left( \operatorname{Id}_{\bs \eta_z^\perp} - \mu H_z \right) \right]^{-1/2}
\label{eq:sorm-prefac-def}
\end{equation}
whenever the matrix inside the determinant is positive definite. In particular,
$
C_{\rm SORM}(z, \lambda_z) = \left[2 I(z) \det G_z  \right]^{-1/2}
$
if evaluated at the Lagrange multiplier from~\eqref{eq:first-order-opt}. LDT shows that the probability mass contributing to the rare event concentrates near $\bs \eta_z$ as $\ve \downarrow 0$. Near the instanton, the normal and tangential displacements have different scales~\cite{tong-vanden-eijnden-stadler:2021,schorlepp-tong-grafke-etal:2023}: the normal displacement is of order~$\ve$, while the tangential displacement is of order~$\sqrt{\ve}$. In the vicinity of~$\bs \eta_z$, we therefore write
\begin{equation}
\bs \eta_{\ve}(s, \bs u) = \bs \eta_z + \ve s \bs e_z + \sqrt{\ve}\bs u, \quad s \in \R, \,\, \bs u \in \bs \eta_z^\perp,
\label{eq:inst-local-coords}
\end{equation}
where $s$ is the parallel coordinate, and $\bs u$ is the tangential coordinate. A Taylor expansion of~$F$ around~$\bs \eta_z$ then gives
\begin{equation*}
F(\bs \eta_\ve) - z = \ve \left[ \frac{r}{\lambda_z}s + \frac{1}{2}\langle \bs u, H_z \bs u \rangle \right] + O\left(\ve^{3/2}\right),
\end{equation*}
so the quadratic approximation of the event condition is $s \geq \tau(\bs u)$, with
\begin{align}
\tau(\bs u) := -\frac{\lambda_z}{2r}\langle \bs u, H_z \bs u \rangle\,.
\label{eq:quad-boundary-def}
\end{align}
Applying this Taylor expansion in~\eqref{eq:tail-prob-def} shows the following sharp asymptotics~\cite{breitung:1984,iltis:1995,tong-vanden-eijnden-stadler:2021,schorlepp-tong-grafke-etal:2023}\footnote{The technical requirements regarding instanton dominance and boundary approximation errors are included in Appendix~\ref{app:proofs-sec3}; we provide a proof of \cref{prop:sharp-asymp} via \cref{prop:vanilla-ibis-moments} (i) below.}:
\begin{proposition}[sharp LDT asymptotics]
\label{prop:sharp-asymp}
Under \cref{assump:standard-ldt}, the rare event probability~\eqref{eq:tail-prob-def} satisfies
    \begin{align}
    P_\ve(z) \overset{\ve \downarrow 0}{\sim} \sqrt{\frac{\ve}{2 \pi}} \; C_{\rm SORM}(z, \lambda_z) \; \exp \left(-\frac{1}{\ve}I(z) \right)\,,
    \end{align}
    where we write $g(\ve) \overset{\ve\downarrow0}{\sim} h(\ve)$ if and only if $\lim_{\ve \downarrow 0} g(\ve)/h(\ve) = 1$.
\end{proposition}
To analyze the variance of IS, we need more stringent assumptions beyond what is necessary for~\cref{prop:sharp-asymp}:
\begin{assumption}[for IS variance] 
\label[assumption]{assump:variance}
In addition to \cref{assump:standard-ldt}, further assume:
\begin{enumerate}[label=(\roman*), leftmargin=*]
    \item that the variance rate function~\cite{bucklew:2013}
    \begin{align}
        \inf_{\bs \eta : F(\bs \eta) \geq z} \frac{1}{2} \lVert \bs \eta + \bs \eta_z \rVert^2
        \label{eq:variance-rate-function}
    \end{align}
    achieves its unique global minimizer at the instanton $\bs \eta_z$.
    \item the stricter curvature condition $\operatorname{Id}_{\bs \eta_z^\perp} - 2 \lambda_z H_z \succ 0$.
\end{enumerate}
\end{assumption}
The additional conditions in \cref{assump:variance} are used in the vanilla IBIS variance and control-variate analysis in \cref{sec:theory-cov}. The localized pecan IBIS variance calculation and the KL calculations in \cref{sec:theory-kl} do not require these additional conditions.
Note that \cref{assump:variance} (i) is a global condition, to guarantee that the instanton $\bs \eta_z$ remains relevant when estimating the variance of the IS proposals~\cite{glasserman-wang:1997,bucklew:2013}. Geometrically, this condition means that $\bs \eta_z$ is the closest point in the event set to $-\bs \eta_z$. \Cref{assump:variance} (ii) is a local condition required for integrability and finiteness of the IS variances, and is, in fact, a standard second-order sufficient optimality condition for~\eqref{eq:variance-rate-function}, as one can easily check. Denoting the eigenvalues of $ \lambda_z H_z$ by $\kappa_i \in \mathbb{R}$, \cref{assump:variance} (ii) requires, for all $i$, $-\infty < \kappa_i < 1/2$, instead of merely $-\infty < \kappa_i < 1$ under \cref{assump:standard-ldt} (iv). Geometrically, the event set's boundary curvatures at $\bs \eta_z$ need to be sufficiently bounded away from the standard normal contours.
\begin{remark}[dominating point condition]
\label[remark]{rem:dominating point}
A standard but strong assumption ensuring that any point $\bs \eta_z$ satisfying the first-order optimality condition \cref{eq:first-order-opt} is the unique global minimizer required by both \cref{assump:standard-ldt} (ii) and \cref{assump:variance} (i) is the \textit{dominating point} (or \textit{half-space}) condition:
\begin{equation}
\{\bs \eta | F(\bs \eta) \geq z\} \subseteq \mathcal{H}(z) = \{ \bs \eta \in \R^N | \langle \bs e_z, \bs \eta - \bs \eta_z \rangle \geq 0 \},
\label{eq:dominating-point}
\end{equation}
such that the event is entirely contained in the half-space $ \mathcal{H}(z) $ with boundary passing through~$\bs \eta_z$ and normal direction $\bs e_z$ along the instanton $\bs \eta_z$~\cite{bucklew:2013}. In particular, this implies that all of the eigenvalues $\kappa_i$ of $\lambda_z H_z$ are non-positive, so that the boundary of the event set is locally convex and does not curve towards the origin in any direction, and hence \cref{assump:variance} (ii) holds as well. Small but positive curvatures $\kappa_i$ are, however, common in practice, and thus we avoid imposing the dominating point condition wherever possible.
\end{remark}

\subsection{The IBIS sampler family} \label{sec:ibis-family}
Now, we specify the instanton-based importance samplers (IBIS) outlined in the introduction in detail. All of them share the same structure: a tangential coordinate $\bs u \sim {\cal N}(\bs 0, \operatorname{Id}_{\bs \eta_z^\perp})$ and a normal coordinate $s$ are drawn, mapped to a sample $\bs \eta$ near
the instanton, and a prefactor variable is evaluated and averaged. The samplers differ only in the law of $s$, in the map $(s, \bs u)\mapsto \bs \eta$,
and in whether a control variable with an easy-to-compute mean
is added. \Cref{tab:ibis-family} summarizes the two proposals, and \cref{alg:ibis-family} gives the common algorithm; the parts printed in \cv{blue} are those added by the control variate, and omitting them yields the plain samplers. The individual methods are described in the subsequent subsections.

\begin{table}[!t]
\caption{Proposals of the IBIS family, as used in \cref{alg:ibis-family}. Each method draws $\bs u \sim {\cal N}(\bs 0, \operatorname{Id}_{\bs \eta_z^\perp})$ and a normal coordinate $s$, and forms $\bs \eta = \bs \eta_z + (\text{normal part}) + (\text{tangential part})$. The row in \cv{blue} is used by the control-variate versions only.}
\label{tab:ibis-family}
\centering
\begin{tabular}{lcc}
\toprule
 & vanilla IBIS & \pecan\ IBIS \\
\midrule
tangential part of $\bs\eta$   & $\sqrt{\ve}\,\bs u$ & $\sqrt{\ve}\,G_z^{-1/2}\bs u$ \\[2pt]
normal part of $\bs\eta$       & $\ve s\,\bs e_z$ & $(\ve s/r)\,\bs e_z$ \\[2pt]
law of $s$ given $\bs u$       & ${\cal N}(0,\ve^{-1})$ & $\mu_\ve^\parallel(\cdot\mid\bs u)$, \eqref{eq:mod-proposal} \\[2pt]
prefactor $W_\ve(s,\bs u)$     & $X_\ve$, \eqref{eq:ibis-def} & $J_\ve$, \eqref{eq:mod-ibis-def} \\[2pt]
deterministic factor $D_\ve$   & $\sqrt{\ve/2\pi}\,e^{-I(z)/\ve}$ & $\sqrt{\ve/2\pi}\,C_{\rm SORM}(z,\lambda_z)\,e^{-I(z)/\ve}$ \\[2pt]
\cv{control variable $V_\ve$, mean $\E[V_\ve]$} & \cv{$Y_\ve$ \eqref{eq:control-variable-def}, $C_{\rm SORM}(z,\lambda_z)$} & \cv{$K_\ve$ \eqref{eq:mod-control-def}, $1$} \\
\bottomrule
\end{tabular}
\end{table}

\begin{algorithm}[!t]
\caption{Generic IBIS estimator (\cv{blue: control-variate steps only})}
\label{alg:ibis-family}
\begin{algorithmic}[1]
\Require noise strength $\ve$, threshold $z$, sample size $M$, instanton $\bs\eta_z$, map $F$, and one column of \cref{tab:ibis-family}\cv{; optionally its control variable $V_\ve$ with known mean $\E[V_\ve]$}
\Ensure estimate $\widehat P_{M,\ve}$ of $P_\ve(z)$
\For{$m=1,\dots,M$}
  \State draw $\bs u^{(m)}\sim{\cal N}(\bs 0,\operatorname{Id}_{\bs\eta_z^\perp})$ and $s^{(m)}$ from the law of $s$ given $\bs u^{(m)}$
  \State $\bs\eta^{(m)}\gets\bs\eta_z+(\text{normal part})+(\text{tangential part})$
  \State $W^{(m)}\gets W_\ve(s^{(m)},\bs u^{(m)})$\cv{, \quad $V^{(m)}\gets V_\ve(s^{(m)},\bs u^{(m)})$}
\EndFor
\State \cv{$\hat c\gets-\widehat{\operatorname{Cov}}_M(W,V)\,/\,\widehat{\operatorname{Var}}_M(V)$} \quad (without control variate: $\hat c\gets 0$)
\State \Return $\displaystyle \widehat P_{M,\ve}\gets D_\ve\,\frac1M\sum_{m=1}^M\Bigl[W^{(m)}\cv{+\hat c\,\bigl(V^{(m)}-\E[V_\ve]\bigr)}\Bigr]$
\end{algorithmic}
\end{algorithm}

\subsubsection{Vanilla IBIS}
\label{sec:vanilla-ibis-def}
The original small-noise distribution is $\pi_\ve = {\cal N}(\bs 0, \ve \operatorname{Id}_N)$\footnote{This is without loss of generality for general Gaussians, whose mean and covariance can be included in $F$.}.
Vanilla IBIS shifts the mean of the Gaussian distribution from the origin to the instanton. The proposal distribution is
$
\mu_\ve = {\cal N}(\bs \eta_z, \ve \operatorname{Id}_N).
$
Since samples are drawn from $\mu_\ve$ instead of $\pi_\ve$, they have to be reweighted by the likelihood ratio
\begin{equation}
\rho_\ve(\bs \eta) := \frac{\mathrm{d}\pi_\ve}{\mathrm{d} \mu_\ve}(\bs \eta) = \exp\left( \frac{1}{\ve}\left( I(z) - \langle \bs \eta, \bs \eta_z \rangle \right) \right)\,.
\label{eq:likelihood}
\end{equation}
Using the local coordinates $\bs \eta = \bs \eta_z + \ve s \bs e_z + \sqrt{\ve}\bs u$ with $s \sim {\cal N}(0, \ve^{-1})$ and $\bs u \sim {\cal N}(0, \operatorname{Id}_{\bs \eta_z^\perp})$, the likelihood ratio becomes
$\rho_\ve(\bs \eta) = \exp\left( -I(z) / \ve - rs \right)$.
This gives the representation
\begin{equation*}
P_\ve(z) = \left( \frac{\ve}{2\pi}\right)^{1/2} e^{-I(z)/\ve} \E[X_\ve]
\end{equation*}
of the tail probability~\eqref{eq:tail-prob-def}, where the vanilla IBIS prefactor estimator is
\begin{equation}
X_\ve = \left( \frac{2\pi}{\ve}\right)^{1/2} \ind_{\{F(\bs \eta_z + \ve s {\bs e}_z + \sqrt{\ve}\bs u) \geq z\}}e^{-rs}.
\label{eq:ibis-def}
\end{equation}
Here, we insert the prefactor $\left( 2\pi / \ve\right)^{1/2}$ in $X_\ve$ on purpose, so that its expectation has an $O(1)$ limit as $\ve \downarrow 0$ in \cref{prop:vanilla-ibis-moments}~(i) below.

\subsubsection{Vanilla IBIS with a control variate}
\label{sec:vanilla-ibis-cv-def}
Vanilla IBIS uses the exact event condition
$
\ind_{\{F(\bs \eta_z + \ve s \bs e_z + \sqrt{\ve}\bs u) \geq z \}}.
$
Near the instanton, the event boundary can be approximated by the quadratic condition
$s \geq \tau(\bs u)$, with $\tau(\bs u)$ given by~\eqref{eq:quad-boundary-def}.
This approximation is used to construct a control variable that is close to $X_\ve$ and has a known expectation. Define
\begin{equation}
Y_\ve := \left( \frac{2\pi}{\ve}\right)^{1/2} \ind_{\{s \geq \tau(\bs u)\}}\exp\left( -rs + \frac{\ve s^2}{2} \right).
\label{eq:control-variable-def}
\end{equation}
The expectation of $Y_\ve$ is known, as we show below in \cref{lem:control-variable-mean}: $\E[Y_\ve] = C_{\rm SORM}(z, \lambda_z)$ is the SORM prefactor in \cref{prop:sharp-asymp}.
We define the control variate prefactor by (see, e.g., \cite{glasserman:2004})
\begin{equation}
X_\ve^{\CV} := X_\ve + c_\ve (Y_\ve - \E[Y_\ve])\,, \quad \text{with optimal coefficient} \quad c_\ve^\ast = -\frac{\cov(X_\ve, Y_\ve)}{\Var(Y_\ve)}.
    \label{eq:cv-coeff-opt}
\end{equation}
In practice, this coefficient is estimated empirically:
$
\hat{c}_{\ve, M} = - \widehat{\cov}_M(X_\ve, Y_\ve) / \widehat{\Var}_M(Y_\ve)
$.
The same samples of $s$ and $\bs u$ are used to compute both $X_\ve$ and $Y_\ve$.

\subsubsection{\Pecan\ IBIS}
\label{sec:mod-ibis-def}

In vanilla IBIS, the normal coordinate $s$ is sampled from a Gaussian distribution with variance $\ve^{-1}$. However, the limiting distribution ``above'' the quadratic boundary (in the orientation of \cref{fig:sampler-sketch}) is a shifted exponential distribution~\cite{schorlepp-tong-grafke-etal:2023}. \Pecan\ IBIS uses this exponential distribution above the quadratic boundary. It also assigns
some probability below the quadratic boundary because the exact event boundary may lie below its quadratic approximation. In the following, we use the curvature-adjusted coordinates that, locally at the instanton, flatten the boundary of the rare-event set:
\begin{equation*}
    \bs \eta_\ve(s,\bs u) = \bs \eta_z + \frac{\ve s}{r}\bs e_z + \sqrt{\ve}G_z^{-1/2} \bs u,
\end{equation*}
where $\bs u \sim {\cal N}(\bs 0, \operatorname{Id}_{\bs \eta_z^\perp})$. Let
\begin{equation*}
b(\bs u) = -\frac{\lambda_z}{2} \left\langle \bs u, G_z^{-1/2} H_z G_z^{-1/2} \bs u \right\rangle 
\end{equation*}
denote the quadratic event boundary in these coordinates. Choose a weight $w_\ve \in (0,1)$, and~$g_\ve$ a probability density on $(0, \infty)$. Conditional on $\bs u$, define the \pecan\ IBIS proposal for $s$ by
\begin{equation}
\mu_\ve^\parallel(s \mid \bs u) = (1-w_\ve)e^{-(s-b(\bs u))}\ind_{\{s \geq b(\bs u)\}} + w_\ve g_\ve(b(\bs u)-s) \ind_{\{s < b(\bs u)\}}.
\label{eq:mod-proposal}
\end{equation}
Thus, (i) with probability $1-w_\ve$, $s = b(\bs u) + E$, $E \sim \text{Exp}(1)$, and (ii)  with probability $w_\ve$, $s = b(\bs u) - V$, $V \sim g_\ve$.
Branch (i) has the exact form of the limiting distribution above the quadratic boundary, and the branch (ii) allows the proposal to sample the part of the exact event that may lie below this asymptotic boundary. Further define
$
h_\ve(s) = \exp\left( -\ve s^2/ ( 2r^2 )\right).
$
The \pecan\ IBIS prefactor variable is
\begin{equation}
J_\ve(s,\bs u) = \ind_{\{F(\bs \eta_\ve(s,\bs u))\geq z\}} h_\ve(s) \begin{cases}
\frac{1}{1-w_\ve}, & s \geq b(\bs u), \\
\frac{e^{-(s-b(\bs u))}}{w_\ve g_\ve(b(\bs u)-s)}, & s < b(\bs u).
\end{cases}
\label{eq:mod-ibis-def}
\end{equation}
The rare-event probability can then be written as
\begin{equation*}
P_\ve(z) = \left( \frac{\ve}{2\pi}\right)^{1/2} C_{\rm SORM}(z, \lambda_z) e^{-I(z)/\ve} \E[J_\ve]\,,
\end{equation*}
where we have effectively factored out the sharp asymptotics from \cref{prop:sharp-asymp}, and are hence only sampling the remaining contributions to the tail probability at finite $\ve$ beyond quadratic terms in $F$. 
A few remarks on this sampler are in order:

\begin{remark}[Design choices and heuristics for \pecan\ IBIS]
\label[remark]{rem:discuss-pecan}
(1) Naive choices for $g_\varepsilon$ are exponential decay (i.e., an exponential distribution) and sub-gaussian decay (half-Gaussian or $\chi(1)$). (2) It is clear that one should take $w_\ve\downarrow0$ as $\ve\downarrow0$, since the second-order approximation is asymptotically sharp. Both choices are revisited in \cref{sec:theory-cov,sec:kdv}.
\end{remark}

To evaluate $G_z^{-1/2} \bs u$ in high dimensions, we typically use Woodbury's formula. Suppose the Hessian $\lambda_z H_z$ has eigenvalues $\kappa_i < 1$, $i = 1, \dots, N-1$, with orthonormal eigenvectors $\bs v_1, \dots, \bs v_{N-1}$. We expect many eigenvalues to satisfy $|\kappa_i| \ll 1$, especially for problems arising from discretizations of well-posed infinite-dimensional problems. Ordering the eigenvalues by decreasing absolute value and truncating after the $k$-th eigenvalue, we set
\begin{equation*}
    G_z^{-1/2} \bs u = \left(\operatorname{Id}_{\bs \eta_z^\perp} - \lambda_z H_z\right)^{-1/2} \bs u \approx \sum_{i=1}^k \frac{1}{\sqrt{1 - \kappa_i}} \langle \bs u, \bs v_i \rangle \bs v_i + \left(\operatorname{Id}_{\bs \eta_z^\perp} - \sum_{i=1}^k \bs v_i^{\otimes 2} \right) \bs u\,.
\end{equation*}

\subsubsection{\Pecan\ IBIS with a control variate}
\label{sec:mod-ibis-cv-def}

The control variate idea can naturally also be applied to the \pecan\ IBIS proposal. The \pecan\ IBIS prefactor $J_\ve$ uses the exact boundary condition and the factor $h_\ve(s)$. Its control variable is obtained by replacing the exact event with the quadratic condition $s \geq b(\bs u)$ and removing $h_\ve(s)$. Define
\begin{equation}
K_\ve(s,\bs u) = \frac{1}{1-w_\ve}\ind_{\{s \geq b(\bs u)\}}.
\label{eq:mod-control-def}
\end{equation}
Since the right branch is selected with probability $1 - w_\ve$, we have
$
\E[K_\ve] = 1
$.
The \pecan\ IBIS control variate prefactor is
\begin{equation*}
J_\ve^{\CV} = J_\ve + d_\ve \left( K_\ve - 1 \right),
\text{ with optimal coefficient }
d_\ve^\ast = -\frac{\cov(J_\ve, K_\ve)}{\Var(K_\ve)}.
\end{equation*}

\section{Coefficient of variation analysis of the different samplers}
\label{sec:theory-cov}

Let $Z_\ve$ denote a single-sample unbiased estimator of the rare-event probability $P_\ve = P_\ve(z)$. Using $M$ independent copies, we define the MC estimator $\hat{P}_{M,\ve}$ as their sample average.
The coefficient of variation (CoV) of $Z_\ve$ is its standard deviation divided by its mean:
\begin{equation}
    \Cov(Z_\ve) = \frac{\sqrt{\Var(Z_\ve)}}{\E[Z_\ve]}.
    \label{eq:cov-def}
\end{equation}
Since the estimator is unbiased, its mean is $P_\ve$. For the average of $M$ independent samples, the variance is reduced by a factor of $M$, and hence the CoV is reduced by a factor of $\sqrt{M}$:
$
    \Cov(\hat{P}_{M,\ve}) = \frac{1}{\sqrt{M}}\Cov(Z_\ve).
$
Thus,
it suffices to study the single-sample estimator asymptotics as $\ve \downarrow 0$. Moreover, since every estimator in \cref{alg:ibis-family} has the form $Z_\ve = D_\ve W_\ve$ with a deterministic factor $D_\ve$, we have $\Cov(Z_\ve) = \Cov(W_\ve)$, and we work with the prefactor variables $X_\ve$ and $J_\ve$ of \cref{tab:ibis-family} throughout. All proofs are deferred to Appendix~\ref{app:proofs-sec3}.

The CoV is the relative root mean squared error of the (unbiased) estimator and hence determines the sample size needed for a prescribed \textit{error level} $\delta>0$:
\begin{equation}
    \Cov(\hat{P}_{M,\ve})^2 = \frac{1}{M}\Cov(Z_\ve)^2 = \delta
    \iff M = \frac{1}{\delta}\Cov(Z_\ve)^2\,. \label{eq:sample-size-cov}
\end{equation}
In general, we are interested in the interaction of the sample size $M$, the noise level $\varepsilon$ (or probability $P_\ve$), and the error level $\delta$. In~\cref{fig:cov-comparison}, $M$ is fixed and we observe the behavior of $\delta$ with respect to $\ve$. 
For instance, it is well-known that for standard MC with $\mathbb{E}[Z_\ve] = P_\ve(z)$ and $\operatorname{Var}(Z_\ve) = P_\ve(z) (1 - P_\ve(z))$, we have $\Cov(\hat{P}_{M,\ve})^2 \overset{\ve \downarrow 0}{\sim} (M P_\ve(z))^{-1}$, and hence $M \overset{\ve \downarrow 0}{\sim} (\delta P_\ve(z))^{-1}$ samples are needed to maintain a constant error $\delta$, i.e., exponentially many by \cref{prop:sharp-asymp}.
We will come back to the question of determining a necessary sample size $M$ given $\ve$ and $\delta$ in \cref{sec:theory-kl}.

\subsection{Vanilla IBIS}
\label{sec:vanilla-ibis-theory}
Recall the vanilla IBIS prefactor $X_\ve$ from~\eqref{eq:ibis-def}, with $s \sim {\cal N}(0, 1/\ve)$ and $\bs u \sim {\cal N}(\bs 0, \operatorname{Id}_{\bs \eta_z^\perp})$ independent.

\begin{proposition}[Mean and variance of vanilla IBIS]
\label{prop:vanilla-ibis-moments}
\begin{enumerate}[label=(\roman*), leftmargin=*]
    \item Under \cref{assump:standard-ldt}, $\E[X_\ve] \to C_{\rm SORM}(z, \lambda_z)$ as $\ve \downarrow 0$, i.e., the mean converges to the SORM prefactor~\eqref{eq:sorm-prefac-def}.
    \item Under \cref{assump:variance},
    $
        \Var(X_\ve) \overset{\ve \downarrow 0}{\sim} \E\left[ X_\ve^2 \right] \overset{\ve \downarrow 0}{\sim} \frac{1}{2}\sqrt{\frac{2\pi}{\ve}}\,C_{\rm SORM}(z, 2\lambda_z).
    $
\end{enumerate}
\end{proposition}
This immediately implies the following:
\begin{corollary}[CoV of vanilla IBIS]
\label{cor:vanilla-ibis-cov}
Under \cref{assump:variance},
\begin{equation}
    \Cov (X_\ve) \overset{\ve \downarrow 0}{\sim} \left( \frac{2\pi}{\ve}\right)^{1/4} \frac{\sqrt{C_{\rm SORM}(z, 2\lambda_z)/2}}{C_{\rm SORM}(z, \lambda_z)}.
    \label{eq:cov-ibis}
\end{equation}
In particular, the CoV diverges like $\ve^{-1/4}$ as $\ve \downarrow 0$, and by~\eqref{eq:sample-size-cov}, maintaining a fixed squared CoV requires (for a fixed $\delta>0$)
\begin{equation}
    M \overset{\ve \downarrow 0}{\sim} \frac{\beta_{\rm CoV}}{\delta}\,\ve^{-1/2}, \qquad \beta_{\rm CoV} := \sqrt{2\pi}\,\frac{C_{\rm SORM}(z, 2\lambda_z)}{2\,C_{\rm SORM}(z, \lambda_z)^2}
    \label{eq:sample-size-ibis}
\end{equation}
samples.
\end{corollary}
The reason for this divergence is a mismatch of scales in the normal direction: the Gaussian proposal spreads $s$ over the width $O(\ve^{-1/2})$, while the weight $e^{-rs}$ confines the contributing samples to an $O(1)$ window above the boundary $\tau(\bs u)$. Only a fraction of order $\sqrt{\ve}$ of the samples therefore carries non-negligible weight, so the effective sample size is $M\sqrt{\ve}$ rather than $M$, consistent with~\eqref{eq:sample-size-ibis}.
This $\ve^{-1/4}$ behavior is indicated by the dashed reference line in \cref{fig:cov-comparison}. We further point out that \eqref{eq:cov-ibis} matches the scaling in~\cite[equation (3-12)]{tong-vanden-eijnden-stadler:2021}, but that it is necessary, as done here, to retain the determinant ratio for sharp asymptotics.

\subsection{Vanilla IBIS with control variate}
\label{sec:vanilla-ibis-cv-general}
The control variable $Y_\ve$ in~\eqref{eq:control-variable-def} differs from $X_\ve$ in two ways.
First, the exact event $\{F \geq z\}$ is replaced by its quadratic approximation $\{s \geq \tau(\bs u)\}$ from~\eqref{eq:quad-boundary-def}. Second, the additional factor $e^{\ve s^2/2}$ cancels the Gaussian density of $s$ when computing the expectation. These choices make the integral over $s$ an exact exponential integral, and the remaining integral over $\bs u$ a Gaussian integral. As shown in \cref{lem:control-variable-mean}, the control variable (by construction) has the known mean 
\[\E[Y_\ve] = C_{\rm SORM}(z, \lambda_z).\]
Near the instanton, the exact and quadratic boundaries agree to leading order, and the additional exponential factor approaches one for bounded $s$. Thus, $X_\ve$ and $Y_\ve$ result in similar contributions in this region, which makes $Y_\ve$ useful as a control variate.

The price for the exact mean is an infinite population variance for every $\ve > 0$ as shown in \cref{prop:control-variable-var} in the appendix. 
This divergence comes from extremely large values of the normal coordinate $s$: the integrand defining the second moment of $Y_\ve$ begins increasing at $s = 2r/\ve$, while typical samples of $s$ have size $1/\sqrt{\ve}$. As the noise decreases, the increasing region moves farther into the Gaussian tail. This explains how an infinite population variance can coexist with the performance seen in finite sample computations in \cref{fig:cov-comparison}. \Cref{fig:cv-sketch} illustrates this separation of scales. If some $\kappa_i>0$, then $\tau(\bs u)$ is unbounded below, and large negative values of $s$ at large $\Vert\bs u\Vert$ cause an additional divergence (\cref{prop:control-variable-var}), requiring absolute values in what follows.

To describe the behavior of finite samples, we thus introduce a cutoff before the increasing region. Fix $0 < a < 2$, set $b_\ve = a r/\ve$, and define the truncated variables
\[X_{\ve,a} := X_\ve \ind_{\{|s| \leq b_\ve\}}, \quad Y_{\ve,a} := Y_\ve \ind_{\{|s| \leq b_\ve\}}. \]
Here, $a \in (0,2)$ needs to be chosen small enough such that, in addition, $\operatorname{Id}_{\bs\eta_z^\perp} - (2 + a/2)\lambda_z H_z \succ 0$, which is possible by \cref{assump:variance}~(ii) and gives Gaussian integrability of remainder terms in the proofs.

By the Gaussian tail bound in \cref{lem:extreme-sample}, all $M$ samples satisfy $|s| \leq b_\ve$ with probability tending to one, e.g., for fixed $M$ as $\ve \downarrow 0$, and the truncated and original variables coincide for every sample. Hence, the empirical means, variances, covariances, and fitted control-variate coefficients computed from them agree. This connects the analysis of the truncated variables to the quantities computed with a finite number of samples.

For the truncated variables, \cref{thm:population-cv-coefficient} shows that the optimal control-variate coefficient satisfies
\[c_{\ve, a}^\ast = -\frac{\cov(X_{\ve, a}, Y_{\ve, a})}{\Var(Y_{\ve, a})} = -1 + O(\sqrt{\ve}). \]
For the coefficient $\hat{c}_{\ve,M}$ fitted from samples, we show convergence to $-1$ under the additional dominating point condition~\eqref{eq:dominating-point} in \cref{thm:empirical-cv-coefficient}; \cref{thm:cv-variance-reduction} below is stated for the population-optimal coefficient.
Thus, to leading order, the control-variate estimator takes the form $X_{\ve, a} - (Y_{\ve, a} - \E[Y_{\ve, a}])$: since the variances of the two truncated variables and their covariance share the same leading term, subtracting $Y_{\ve, a}$ cancels the leading contribution to the variance, while adding back its expectation preserves the mean of $X_{\ve, a}$.

To understand the variance that remains, we examine the difference between the exact event boundary and its quadratic approximation. For a given tangential coordinate $\bs u$, let $s_\ve^\ast(\bs u)$ denote the normal coordinate of the exact boundary, which has expansion
\[s_{\ve}^\ast (\bs u) = \tau(\bs u) + \sqrt{\ve} c_3(\bs u) + O(\ve), \]
uniformly on bounded tangential regions, where $c_3$ is the cubic boundary correction derived in Appendix~\ref{app:control-variate-asymptotics}. More precisely, by \cref{lem:boundary-expansion}, the remainder is bounded by $C\ve(1 + \Vert \bs u\Vert)^4$ on the growing region $\Vert \bs u \Vert \leq \delta_0/\sqrt{\ve}$, which is what allows us to state the following results for the full Gaussian law of $\bs u$, see \cref{prop:cv-moments}.
Between $s = \tau(\bs u)$ and $s = s_\ve^\ast(\bs u)$, one event indicator equals one and the other equals zero, so their contributions do not cancel. The width of this region is $\sqrt{\ve}|c_3(\bs u)|$ to leading order. This boundary mismatch determines the leading variance in the following proposition. To state the variance reduction, define the weight $q(\bs u):= e^{-2r \tau(\bs u)} = e^{\lambda_z \langle \bs u, H_z \bs u\rangle}$.

\begin{proposition}[Variance reduction by truncated control variate]
\label{thm:cv-variance-reduction}
Suppose \cref{assump:variance} holds, and let $a$ be as above. Define
\[X_{\ve, a}^{\CV} := X_{\ve, a} + c_{\ve, a}^\ast (Y_{\ve, a} - \E[Y_{\ve, a}]), \]
where $c_{\ve, a}^\ast$ minimizes the variance. Then
\begin{equation*}
\frac{\Var(X_{\ve,a}^{\CV})}{\Var(X_{\ve,a})} = C_{\CV}\sqrt{\ve} + O(\ve), \qquad C_{\CV} := 2r\,\frac{\E[q(\bs u)\,|c_3(\bs u)|]}{\E[q(\bs u)]},
\end{equation*}
and consequently
\begin{equation*}
    \operatorname{CoV}(X_{\ve,a}^{\CV}) \overset{\ve \downarrow 0 }{\longrightarrow} \left(\frac{\pi}{2}\right)^{1/4}\frac{\sqrt{C_{\CV}\, C_{\rm SORM}(z, 2 \lambda_z)}}{C_{\rm SORM}(z, \lambda_z)}\,.
\end{equation*}
By~\eqref{eq:sample-size-cov}, the sample size required for a fixed squared CoV 
therefore stays bounded as $\ve \downarrow 0$, in contrast to~\eqref{eq:sample-size-ibis}.
\end{proposition}

\subsection{\Pecan\ IBIS with a scaled distribution below the parabola}
\label{sec:mod-ibis-general}
\begin{figure}
    \centering
    \includegraphics[width=0.8\textwidth]{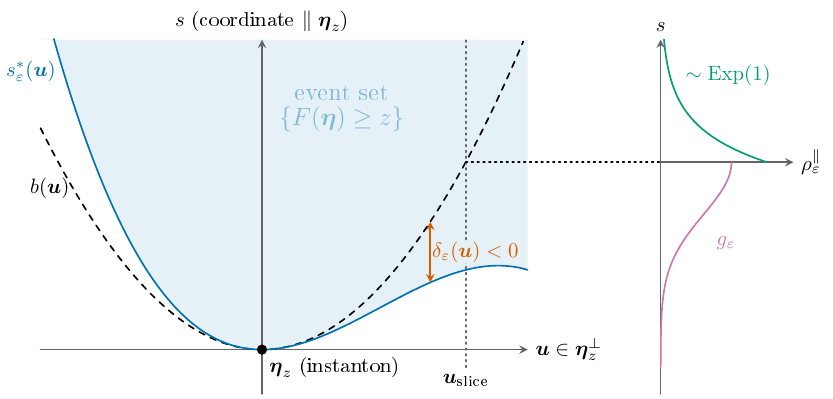}
    \caption{Sketch of the local event boundary geometry near the instanton $\boldsymbol{\eta}_z$. The true boundary $s_\varepsilon^*(\boldsymbol{u})$ may deviate from the quadratic approximation $b(\boldsymbol{u})$, generating regions where $\delta_\varepsilon(\boldsymbol{u}) < 0$. \Pecan\ IBIS samples above $b(\boldsymbol{u})$ using an exponential tail, and captures the region below using another distribution $g_\varepsilon$, as shown in the cross-section along $\boldsymbol{u}_{\text{slice}}$. 
    }
    \label{fig:mod-ibis-sketch}
\end{figure}

We now show that the branch of \pecan\ IBIS below the parabola (see \cref{fig:mod-ibis-sketch}) is effective for a broad class of scaled densities $g_\ve$; no specially optimized distribution is required.

Recall the proposal and prefactor $J_\ve$ introduced in \cref{sec:mod-ibis-def}. For a fixed tangential coordinate $\bs u$, let $s^\ast_\ve(\bs u)$ denote the exact event boundary and let $b(\bs u)$ be its quadratic approximation. We write $\delta_\ve(\bs u) = s^\ast_\ve(\bs u) - b(\bs u)$ for the boundary discrepancy. Thus, negative values of $\delta_\ve$ identify the region that must be sampled by the branch below the parabolic boundary approximation. 

We fix $R > 0$ and restrict the analysis to 
\[\mathcal{B}_R := \{\bs u: \Vert \bs u \Vert \leq R\}.\]
Let $\E_R$, $\Var_R$, and $\cov_R$ denote expectation, variance, and covariance conditioned on $\bs u \in \mathcal{B}_R$. The results in this subsection are stated under this conditional law, with $R$ fixed as $\ve \downarrow 0$, since the density below the parabola must cover the range of the boundary discrepancy, which is bounded only on $\mathcal{B}_R$\footnote{The restriction $\bs u\in\mathcal{B}_R$ is satisfied by all $M$ samples with high probability as long as $M$ is not too large, concretely $R \gtrsim \sqrt{N-1} + \sqrt{2 \log M}$ by a standard chi distribution tail bound and the union bound.
In contrast to \cref{sec:vanilla-ibis-cv-general}, this condition does not become weaker as $\ve \downarrow 0$. Moreover, the constants in \cref{thm:scaled-left-proposal} below depend on such an $R$ and may be large, see the discussion after that proposition.\label{fn:R-tube}}.
As in \cref{sec:vanilla-ibis-cv-general}, \cref{lem:boundary-expansion} yields the expansion $\delta_\ve(\bs u) = \sqrt{\ve} c_3(\bs u) + O(\ve)$ uniformly on $\mathcal{B}_R$, where $c_3$ now denotes the odd cubic boundary correction in the curvature-adjusted coordinates of \cref{sec:mod-ibis-def} (it differs from the $c_3$ of \cref{sec:vanilla-ibis-cv-general} by this change of variables). Since both the Gaussian distribution and $\mathcal{B}_R$ are symmetric, $\E_R[c_3(\bs u)] = 0$. Define $m_1 = \E_R[(c_3(\bs u))_+] = \E_R[(c_3(\bs u))_-]$. 
Since $\mathcal{B}_R$ does not restrict the normal coordinate, we also truncate it, as in \cref{sec:vanilla-ibis-cv-general}: fix $0 < \omega \leq \delta_0$, with $\delta_0$ from \cref{lem:boundary-expansion} in the curvature-adjusted coordinates, and set $J_{\ve,\omega} := J_\ve \ind_{\{|\ve s| \leq \omega\}}$. For $\bs u \in \mathcal{B}_R$ and $\ve$ small, the event within this cutoff is then exactly $\{s \geq s_\ve^\ast(\bs u)\}$ by \cref{lem:boundary-expansion}. The restriction on $M$ through this $s$-cutoff does become weaker as $\ve \downarrow 0$, in analogy to \cref{sec:vanilla-ibis-cv-general}.
The calculation in Appendix~\ref{app:modified-ibis-moments} gives the basic expansion
\begin{equation*}
\Var_R(J_{\ve,\omega}) = w_\ve + \frac{A_\ve(g_\ve)}{w_\ve} - m_1 \sqrt{\ve} + \text{higher order terms,}
\end{equation*}
where
$
A_\ve(g_\ve) = \E_R\big[ \int_0^{(\delta_\ve(\bs u))_-} \frac{e^{2v}}{g_\ve(v)}\,\mathrm{d}v \big].
$
The first term is the price of diverting mass $w_\ve$ below the parabola, the second the cost of covering the region where the true boundary lies below it with a density of total mass only $w_\ve$, and the third the gain from the part of the exact event that lies above the parabola; balancing the first two terms is what yields $w_\ve \sim \sqrt{\ve}$.
We consider densities below the parabola that are scaled to the $\sqrt{\ve}$-width of the boundary discrepancy: for a fixed probability density $g$ on $(0, \infty)$, continuous and strictly positive on an interval containing the range of $(c_3(\bs u))_-$ for $\bs u \in \mathcal{B}_R$, let
\begin{equation}
g_\ve(v) := \ve^{-1/2} g\left( \ve^{-1/2} v\right), \qquad D_g := \E_R\bigg[ \int_0^{(c_3(\bs u))_-} \frac{1}{g(x)}\,\mathrm{d}x\bigg].
\label{eq:scaled-left-density}
\end{equation}
The next result gives the $\Cov$ in $\mathcal B_R$ of \pecan\ IBIS with this scaled density below the quadratic boundary.

\begin{proposition}[Localized variance for general scaled proposals]
\label{thm:scaled-left-proposal}
Suppose \cref{assump:standard-ldt} holds, fix $R > 0$, 
let $g_\ve$ and $D_g$ be as in \eqref{eq:scaled-left-density}, and let the bottom-branch probability be $w_\ve = \alpha \sqrt{\ve}$ for some $\alpha > 0$. Then
\begin{equation*}
\Var_R(J_{\ve,\omega}) = \left(\alpha + \frac{D_g}{\alpha} - m_1\right) \sqrt{\ve} + o(\sqrt{\ve}).
\end{equation*}
The leading constant is positive. Consequently,
\begin{equation*}
\Cov_R(J_{\ve,\omega}) = \left(\alpha + \frac{D_g}{\alpha} - m_1\right)^{1/2} \ve^{1/4} + o(\ve^{1/4}).
\end{equation*}
\end{proposition}

The restriction to $\mathcal{B}_R$ is essential and not merely technical. For the light-tailed densities of \cref{rem:discuss-pecan}, e.g., $g(x)=\beta e^{-\beta x}$, $1/g$ grows exponentially while $(c_3)_-$ is cubic in $\bs u$, so that $D_g=\infty$ for $R=\infty$ 
(see Appendix~\ref{app:modified-ibis-moments}). Heavy-tailed densities $g$, e.g., with power-law tails, keep $D_g$ finite for $R=\infty$, but passing to the limit $R\to\infty$ in \cref{thm:scaled-left-proposal} would additionally require global conditions on the event set.

Note that the leading variance in \cref{thm:scaled-left-proposal} is minimized at $\alpha = \sqrt{D_g}$. Hence, the asymptotically optimal mixture weight within this scaled family is $w_\ve^\ast \sim \sqrt{D_g \; \ve}$, the corresponding variance satisfies $\Var_R({J_{\ve,\omega}}) \sim (2\sqrt{D_g} - m_1) \sqrt{\ve}$, and any sufficiently regular density scaled to the natural $\sqrt{\ve}$-boundary-discrepancy scale yields a variance of order~$\sqrt{\ve}$.
Although the exact constants may be intractable, \cref{thm:scaled-left-proposal} suggests choosing $w_\ve=\alpha\sqrt{\ve}$ for constant $\alpha$, while ensuring $w_\ve\in(0,1)$ for all $\ve$. In our experiments we hence use $w_\ve=\min\{\alpha\sqrt{\ve},1/2\}$, so that for large $\ve$ (an “uninformative” case) the distributions above and below the parabola receive equal weight.

\begin{remark}
\label[remark]{rem:unscaled-left-proposal}
If the density $g_\ve=g$ is chosen to not depend on $\ve$ (and is continuous and positive at zero), the variance converges more slowly. In this case, taking $w_\ve = \alpha \ve^{1/4}$ gives
\begin{equation*}
\Var_R(J_{\ve,\omega}) = \left(\alpha + \frac{m_1}{\alpha g(0)} \right)\ve^{1/4} + o(\ve^{1/4}).
\end{equation*}
The best leading constant is obtained when $\alpha = \sqrt{m_1 / g(0)}$. Thus, an unscaled proposal below the parabola still produces vanishing variance, but scaling the proposal with $\sqrt{\ve}$ improves the variance rate from $\ve^{1/4}$ to $\ve^{1/2}$. 
\end{remark}

\begin{remark}
\label[remark]{rem:optimal-left-distribution}
The variance expansion also identifies a formal optimum. Let $S_\ve(v) = \PP_R(\delta_\ve(\bs u) < -v)$ and ${\cal Z}_\ve = \int_0^\infty e^v \sqrt{S_\ve(v)}\,\mathrm{d}v$. Minimizing the leading variance over all densities gives $g_\ve^\ast(v) =  e^v \sqrt{S_\ve(v)} / {\cal Z}_\ve$, with corresponding mixture weight $w_\ve^\ast = {\cal Z}_\ve$. Since $S_\ve$ depends on the unknown exact event boundary, this distribution is generally intractable in practice. \Cref{thm:scaled-left-proposal} is more useful computationally, showing that the $\sqrt{\ve}$ variance rate holds for a broad class of readily chosen scaled densities.
Appendix \ref{app:toy-example} presents a two-dimensional example in which the optimal density can be evaluated and sampled numerically, and compares its shape and sampling performance with simpler bottom-branch proposals.
\end{remark}

\section{Kullback--Leibler divergence analysis of the different samplers}
\label{sec:theory-kl}
Complementary to the CoV analysis, we now investigate the proposed samplers within the framework of \cite{chatterjee-diaconis:2018}. Their approach splits the question of an appropriate sample size $M$ for IS into a necessary and a sufficient value of $M$, with a phase transition at a \textit{critical sample size}. Although the ideas in \cite{chatterjee-diaconis:2018,sanz-alonso:2018} have become guiding heuristics, sharp evaluation of concrete samplers against this criterion remains uncommon. The explicit small-noise asymptotics available here enable such an evaluation from the information-centric perspective.
Recall that $\pi_\ve = {\cal N}(\bs 0, \ve \operatorname{Id})$ is the original parameter distribution, that $\mu_\ve$ is any of the IS proposals, and we write $\rho_\ve(\bs\eta)=\frac{\mathrm{d} \pi_\ve}{\mathrm{d}  \mu_\ve}(\bs \eta)$ as in~\eqref{eq:likelihood}. Samples from~$\mu_\ve$ are used to estimate the event probability via $M$ iid copies of~$Z_\ve$.
In this section, we make it explicit through subscripts which measure is used when taking expectations, unless it is clear from context.
The KL divergence is defined as $\mathrm{KL}(\pi_\ve\|\mu_\ve)=\mathbb{E}_{\pi_\ve}[\log \rho_\ve(\bs \eta)]$, and the $\chi^2$ divergence is $\chi^2(\pi_\ve\|\mu_\ve)=\mathbb{E}_{\pi_\ve}[\rho_\ve(\bs \eta)]-1$. 
The optimal, but generally intractable IS proposal (the ``target'' below) is the conditional distribution $\mathrm{d} \pi_{\varepsilon,z}(\bs\eta) := \frac{\mathrm{d}\pi_{\varepsilon}(\bs\eta)}{P_\ve(z)}\mathbbm{1}_{\{F(\bs\eta) \geq z\}}$. Our analysis of the CoV in \cref{sec:theory-cov} then relates to the $\chi^2$-divergence via (see, e.g., \cite{sanz-alonso:2018})
\begin{equation*}
     \Cov(\hat{P}_{M,\ve})^2=\frac{1}{MP_\ve(z)^2}\mathrm{MSE}(Z_\ve)=\frac{1}{M}\chi^2(\pi_{\varepsilon,z}\|\mu_\ve)\,.
\end{equation*}
An analysis in the $\chi^2$-divergence (equivalently, the CoV or MSE) 
involves second moments and is therefore 
sensitive to outliers, causing rare, very large weights to dominate.  The analysis in \cite{chatterjee-diaconis:2018}, based on the KL divergence of the target $\pi_{\ve, z}$ from the proposal $\mu_\ve$, instead
averages the logarithm of the weights, which dampens the influence of rare, large weights.

For example, 
let $\pi = \mathcal{N}(0,1)$, $\mu = \mathcal{N}(0,1/2)$, and consider the event $\{\eta\in\R:\ \eta\geq z\}$ with probability $P(z) := \pi(\{\eta \geq z\})$ and conditional law $\pi_z$. Then
\begin{align*}
    \chi^2(\pi_z\|\mu)&=\frac{1}{2\sqrt{\pi}\,P(z)^2}\int_z^\infty1\,\mathrm{d}\eta-1=\infty\,,\\
     \mathrm{KL}(\pi_z\|\mu)&=-\log P(z)-\frac{1}{2}\log2+\frac{1}{2}\mathbb{E}_{\pi_z}[\eta^2]<\infty\,.
\end{align*}
Here, the MSE of the IS estimator is infinite, yet the KL divergence is finite and \cref{thm:chatter-diac} below still applies; in general, $\exp\{\mathrm{KL}(\pi_z\|\mu)\}\leq\chi^2(\pi_z\|\mu)+1$ by Jensen's inequality (see, e.g., \cite{gibbs-su:2002}), with an unbounded gap.

\subsection{Sufficient and necessary sample size}

We start by restating the main result of \cite{chatterjee-diaconis:2018} for the rare event sampling problem at hand. 
We denote $\rho_{\ve,z}(\bs\eta):= \frac{\mathrm{d}\pi_{\varepsilon,z}}{\mathrm{d}\mu_\ve}(\bs \eta)$ below. Then, \cite[Theorem 1.3]{chatterjee-diaconis:2018} states the following\footnote{Specialized here for integrand $f\equiv1$, for which the square root in their concentration term can be dropped, cf.\ Appendix~\ref{app:proofs-sec4}, and with $t$ rescaled.}:
\begin{theorem}[Required sample size in IS for rare events, \cite{chatterjee-diaconis:2018}]
\label{thm:chatter-diac}
    \begin{enumerate}[label=(\roman*), leftmargin=*]
        \item (``$M$ large enough'') If $M=\exp\{ \mathrm{KL}(\pi_{\ve,z}\|\mu_\ve)+2t\}$ for a $t>0$, then
    \begin{equation}
        \frac{\mathbb{E}\left[\left|\hat{P}_{M,\ve}(z)-P_\ve(z)\right|\right]}{\mathbb{E}[\hat{P}_{M,\ve}(z)]}\leq e^{-t/2}+2\mathbb{P}_{\pi_{\ve,z}}\Big[\log\rho_{\ve,z}(\bs\eta)>\mathrm{KL}(\pi_{\ve,z}\|\mu_\ve)+t\Big]\,.
        \label{eq:cd-suff}
    \end{equation}
        \item (``$M$ too small'') If $M=\exp\{\mathrm{KL}(\pi_{\ve,z}\|\mu_\ve)-2t\}$ for a $t>0$, then for any $\zeta \in (0,1)$:
     \begin{equation}
         \mathbb{P}\left[\frac{\hat{P}_{M,\ve}(z)}{P_\ve(z)}\geq\zeta\right]\leq e^{-t}+\frac{1}{\zeta}\mathbb{P}_{\pi_{\ve,z}}\Big[\log\rho_{\ve,z}(\bs\eta)\leq\mathrm{KL}(\pi_{\ve,z}\|\mu_\ve)-t\Big]\,.
         \label{eq:cd-nece}
     \end{equation}
    \end{enumerate}
\end{theorem}
Part (i) bounds the normalized absolute error by a term that decays as soon as $M$ exceeds $e^{\mathrm{KL}(\pi_{\ve,z}\|\mu_\ve)}$, plus the probability that the log-weight exceeds its mean by more than $t$. The necessary condition (ii) shows that for $M$ below the threshold, the estimator is, with high probability, well below its mean $P_\ve(z)$: with too few samples, IS systematically underestimates the probability because the weights that carry most of the mass have not yet been sampled.
Taken together, \cref{thm:chatter-diac} therefore identifies the exponential of the KL divergence $M = \exp \{\operatorname{KL}(\pi_{\ve, z} \| \mu_\ve) \}$ as the \textit{critical}
threshold for the sample size.

The error measure in (i) is bounded by the CoV studied in \cref{sec:theory-cov}: by Cauchy--Schwarz,
\begin{align}
    \frac{\mathbb{E}\left[\left|\hat{P}_{M,\ve}(z)-P_\ve(z)\right|\right]}{\mathbb{E}[\hat{P}_{M,\ve}(z)]}&\leq\Cov(\hat{P}_{M,\ve})=\frac{1}{\sqrt{M}}\sqrt{\chi^2(\pi_{\varepsilon,z}\|\mu_\ve)}\,.
    \label{eq:chi2-bound}
\end{align}
Thus, (i) replaces $\chi^2(\pi_{\ve,z}\|\mu_\ve)$ by $e^{\mathrm{KL}(\pi_{\ve,z}\|\mu_\ve)}\le\chi^2(\pi_{\ve,z}\|\mu_\ve)+1$, which may be finite when $\chi^2$ is not, at the price of the concentration term and a slower decay in $M$ (see \cref{rem:error-rate-M} below).
Both parts of the theorem are informative only if $\log\rho_{\ve,z}$ concentrates around its mean under $\pi_{\ve,z}$. Cantelli's inequality bounds the concentration terms in~\eqref{eq:cd-suff} and~\eqref{eq:cd-nece} by ${\Var_{\pi_{\ve,z}}\left(\log\rho_{\ve,z}(\bs\eta)\right)}/{(\Var_{\pi_{\ve,z}}\left(\log\rho_{\ve,z}(\bs\eta)\right)+t^2)}$, so the sharpness of the threshold is governed by the variance of the log importance weight, which we compute for our samplers below.

\begin{remark}[Error rate in sample size $M$]
\label[remark]{rem:error-rate-M}
The bound in \cref{thm:chatter-diac} (i) does not recover the canonical $O(M^{-1/2})$ Monte Carlo error rate. Because the theorem couples the sample size $M$ and the parameter $t$ via $M = \exp\{\mathrm{KL}(\pi_{\ve,z}\|\mu_\ve)+2t\}$, solving for $t$ and substituting it into the first term $e^{-t/2}$ on the RHS of \eqref{eq:cd-suff} yields an explicit error decay of $O(M^{-1/4})$. While the $\chi^2$-based bound in~\eqref{eq:chi2-bound} formally yields an $O(M^{-1/2})$ rate, it may be unbounded in rare-event settings (as demonstrated in our earlier example), rendering it empty.
\end{remark}

\subsection{Direct MC}
As a reference, we first apply \cref{thm:chatter-diac} to the direct MC proposal $\mu_\ve = \pi_\ve$. In this case, the target density ratio is constant on the event: $\rho_{\ve,z}(\bs\eta)=1/P_\ve(z)$ when $F(\bs\eta)\geq z$, and hence $\log\rho_{\ve,z}(\bs\eta)=-\log P_\ve(z)$ $\pi_{\ve,z}$-almost surely. Thus, $\mathrm{KL}(\pi_{\ve,z}\|\mu_\ve)=-\log P_\ve(z)$ and $\Var_{\pi_{\ve,z}}(\log\rho_{\ve,z}(\bs\eta))=0$. The log-weight does not fluctuate at all under the target, so the concentration terms in \cref{thm:chatter-diac} vanish. Its two parts reduce to
\begin{align*}
    \text{(i)}\quad\frac{\mathbb{E}\left[\left|\hat{P}_{M,\ve}(z)-P_\ve(z)\right|\right]}{\mathbb{E}[\hat{P}_{M,\ve}(z)]}\leq e^{-t/2}
    \;\text{ and }\;
    \text{(ii)}\quad\mathbb{P}\left[\frac{\hat{P}_{M,\ve}(z)}{P_\ve(z)}\geq\zeta\right]\leq e^{-t}\,,
\end{align*}
for $M = e^{\pm2t} / P_\ve(z)$, $t > 0$. That is, with a constant multiple more than $1/P_\ve(z)$ samples, the relative error is small, and with a constant multiple fewer, the estimator is, with high probability, far below the true value. The critical sample size is therefore $M = 1/P_\ve(z)$, the expected number of samples needed to observe a single event. This matches the CoV-based analysis in \cref{sec:theory-cov}, which also requires $M \propto 1/P_\ve(z)$; by \cref{prop:sharp-asymp}, the critical sample size grows exponentially as $\ve \downarrow 0$.

\subsection{Vanilla IBIS}
We now adjust the analysis of \cref{sec:vanilla-ibis-theory} to the KL framework. All proofs are available in Appendix~\ref{app:proofs-sec4}.
\begin{proposition}[KL divergence of vanilla IBIS]
\label{prop:vanilla-ibis-KL}
Under \cref{assump:standard-ldt}, we have
\begin{align*}
    \mathrm{KL}(\pi_{\ve,z}\|\mu_\ve) &= \tfrac{1}{2}\log(1/\ve)+C_{\rm KL}(z, \lambda_z)+o(1),\quad\text{with}\\[1ex]
    C_{\rm KL}(z, \lambda_z)&:=\tfrac{1}{2}\log2\pi-\log C_{\rm SORM}(z, \lambda_z) + \tfrac{1}{2}\operatorname{tr}\left(\lambda_zH_zG^{-1}_z\right) - 1\,.
\end{align*}
In particular, the KL divergence grows like $\frac12 \log(1/\ve)$ as $\ve \downarrow 0$. This implies the critical sample size $M=\exp\{ \mathrm{KL}(\pi_{\ve,z}\|\mu_\ve)\} = \ve^{-1/2}e^{C_{\rm KL}(z, \lambda_z)}\,(1+o(1))$. Moreover, the variance of the log importance weight satisfies
\begin{equation*}
    \Var_{\pi_{\ve,z}}\left(\log\rho_{\ve,z}(\bs\eta)\right)\longrightarrow 1+\tfrac{1}{2}\operatorname{tr}\left(\left(\lambda_zH_zG_z^{-1}\right)^2\right)\quad\text{as }\ve\downarrow0.
\end{equation*}
\end{proposition}
The proof is a direct computation in instanton-adapted coordinates: writing $\mathrm{KL}(\pi_{\ve,z}\|\mu_\ve)=-\log P_\ve(z)-I(z)/\ve-r\,\E_{\pi_{\ve,z}}[s]$, the first two terms are the negative $\log$ of the SORM prefactor,
which produces the $\frac12\log(1/\ve)$ growth, while the mean normal coordinate $s$ under the conditional law contributes $-1$ from the exponential profile along $\bs e_z$ and $\frac12\operatorname{tr}(\lambda_zH_zG_z^{-1})$ from the curvature of the event boundary. 
The same approach gives the two terms in the limiting variance of the log weight. Since this limit is finite and independent of $\ve$, Cantelli's inequality bounds the concentration terms in \cref{thm:chatter-diac} uniformly in $\ve$; the threshold $M=e^{\mathrm{KL}(\pi_{\ve,z}\|\mu_\ve)}$ is thus sharp up to $\ve$-independent factors $e^{\pm2t}$.

The corresponding $\ve$-scaling of the critical sample size of vanilla IBIS, $M \overset{\ve \downarrow 0}{\sim} \beta \times \ve^{-1/2}$ for $\beta$ constant, matches the CoV-based computations in \cref{cor:vanilla-ibis-cov}. However, the constants are different, namely
$
  \beta_{\rm KL} = \sqrt{2 \pi} C_{\rm SORM}(z, \lambda_z)^{-1} \exp\left(\frac{1}{2}\operatorname{tr}\left(\lambda_zH_zG^{-1}_z\right) - 1\right)
$
for the KL-based analysis, and
$
    \beta_{\rm CoV} = \sqrt{2 \pi} C_{\rm SORM}(z, 2 \lambda_z) / (2 C_{\rm SORM}(z, \lambda_z)^2)
$
for the CoV-based one. Note that $\beta_{\rm CoV}$ involves $C_{\rm SORM}(z,2\lambda_z)$ and thus requires \cref{assump:variance}(ii), whereas $\beta_{\rm KL}$ only involves $G_z$ for which
\cref{assump:standard-ldt} suffices. Furthermore, the following holds.

\begin{lemma}[Constant ratio of vanilla IBIS]
\label{lem:vanilla-ibis-comp}
Under \cref{assump:variance}, we have
\begin{equation*}
    \frac{\beta_{\rm KL}}{\beta_{\rm CoV}}=\frac{2}{e}\left(\prod_{i=1}^{N-1}(1-\xi_i)e^{\xi_i}\right)^{\frac{1}{2}}\leq\frac{2}{e}
\end{equation*}
for $\xi_i\coloneqq\frac{\kappa_i}{1-\kappa_i}$, where $\kappa_i$ are the eigenvalues of $\lambda_z H_z$.
\end{lemma}
The term $\prod_{i=1}^{N-1}(1-\xi_i)e^{\xi_i}$ is the Carleman--Fredholm determinant $\det_2(\operatorname{Id}_{\bs \eta_z^\perp}-M_z)$ of $M_z\coloneqq\lambda_zH_zG^{-1}_z$ \cite{schorlepp-grafke:2025}. Thus, the two analyses agree up to an order-one factor: for a flat boundary ($\kappa_i=0$), the KL-critical sample size is exactly $2/e$ times the CoV-based one at error level $\delta=1$ in~\eqref{eq:sample-size-ibis}. The gap widens with curvature: as some $\kappa_i\uparrow 1/2$, the limit allowed for in \cref{assump:variance}(ii), $C_{\rm SORM}(z,2\lambda_z)$ and hence $\beta_{\rm CoV}$ diverge, while $\beta_{\rm KL}$ stays finite, i.e., the KL analysis does not see the large weights that inflate the variance. In the opposite direction, $\kappa_i\to-\infty$, both constants diverge while their ratio stays bounded; i.e., a strongly convex event set is sampled inefficiently by vanilla IBIS, and both analyses agree on this.

\subsection{\Pecan\ IBIS}

In contrast to the CoV analysis in \cref{sec:mod-ibis-general}, no restriction to $\mathcal{B}_R$ is needed here, since for a large class of densities $g$, the log-weights grow at most polynomially away from the instanton. Thus, the region where the boundary expansion of \cref{lem:boundary-expansion} fails contributes negligibly to the KL divergence. The proof is again available in Appendix~\ref{app:proofs-sec4}, and we find:

\begin{proposition}[KL divergence of \pecan\ IBIS]
\label{prop:mod-ibis-KL}
Suppose \cref{assump:standard-ldt} holds, let $g_\ve(v)=\ve^{-1/2}g(\ve^{-1/2}v)$ for a probability density $g$ on $(0,\infty)$ with $|\log g(x)|\leq C_g(1+x)^p$ for all $x>0$ and some constants $C_g,p\geq0$ (e.g., exponential or half-normal $g$), and let $w_\ve=\alpha\sqrt{\ve}$ for some $\alpha>0$. Define
\begin{equation*}
    m_1:=\E\big[(c_3(\bs u))_-\big],\qquad\widetilde{D}_g\coloneqq\mathbb{E}\left[\int_0^{(c_3(\bs u))_-}\log\left(\frac{1}{g(x)}\right)\mathrm{d}x\right],
\end{equation*}
with $c_3$ in the curvature-adjusted coordinates as in \cref{sec:mod-ibis-general}; these constants are finite and are the limits as $R\to\infty$ of their counterparts restricted to $\mathcal{B}_R$. Then
\begin{equation*}
    \mathrm{KL}(\pi_{\ve,z}\|\mu_\ve) = \sqrt{\ve}\left(\alpha+\widetilde{D}_g-m_1\log\alpha\right)+o(\sqrt{\ve})\,.
\end{equation*}
Moreover, the variance of the log importance weight satisfies
\begin{equation*}
    \Var_{\pi_{\ve,z}}\left(\log\rho_{\ve,z}(\bs\eta)\right)=\sqrt{\ve}\;\E\left[\int_0^{(c_3(\bs u))_-}\left(\log\frac{1}{g(x)}-\log\alpha\right)^2\mathrm{d}x\right]+o(\sqrt{\ve})\,.
\end{equation*}
\end{proposition}

The KL divergence vanishes like $\sqrt{\ve}$, and the critical sample size $M=
\exp\{ \mathrm{KL}(\pi_{\ve,z}\|\mu_\ve)\} =
1+\sqrt{\ve}\big(\alpha+\widetilde{D}_g-m_1\log\alpha\big)+o(\sqrt{\ve})$ tends to one. Its leading constant is minimized at $\alpha = m_1$, i.e., at $w^*_\ve=m_1\sqrt{\ve}$, for which the proposal assigns, to leading order, the same mass $m_1\sqrt{\ve}$ to the region below the parabola as the target. Since $e^{\mathrm{KL}(\pi_{\ve,z}\|\mu_\ve)}\to1$, the necessary condition in \cref{thm:chatter-diac}~(ii) is void, and the relevant statement is the sufficient condition~(i). As the variance of the log weight in \cref{prop:mod-ibis-KL} is $O(\sqrt{\ve})$, Cantelli's inequality shows that $M=e^{2t}(1+O(\sqrt{\ve}))$ samples yield a normalized error of at most $e^{-t/2}+O(\sqrt{\ve})/t^2$.

The KL analysis and the CoV analysis in \cref{thm:scaled-left-proposal} both give a correction of order $\sqrt{\ve}$, where the former notably does not require restriction to $\mathcal{B}_R$.
By~\eqref{eq:sample-size-cov}, the CoV-based sample size is $M\sim\beta_{\rm CoV}\sqrt{\ve}/\delta$, while the KL-critical one is $M=1+\beta_{\rm KL}\sqrt{\ve}+o(\sqrt{\ve})$. In both, the $\ve$-dependent part is of order $\sqrt{\ve}$, i.e., a bounded number of samples suffices as $\ve\downarrow0$. The constants for \pecan\ IBIS are
\begin{equation*}
    \beta_{\rm KL} = m_1+\widetilde{D}_g-m_1\log m_1\qquad\text{and}\qquad \beta_{\rm CoV} = 2\sqrt{D_g} - m_1 \; (\text{on } {\cal B}_R)\,.
\end{equation*}
Here, $\widetilde{D}_g$ is less restrictive than $D_g$, since its finiteness only requires $\log(1/g)$, rather than $1/g$, to be integrable on the range of $(c_3(\bs u))_-$; for the light-tailed densities of \cref{rem:discuss-pecan}, $\widetilde{D}_g$ is finite for $R=\infty$, whereas $D_g$ is not. Thus, as in the introductory example of this section, $\widetilde{D}_g$ and the ``critical'' sample size can remain finite when $D_g$, and hence the variance, is infinite. Moreover, the two analyses give different, explicit recipes for the bottom-branch weight: $w_\ve^\ast = m_1\sqrt{\ve}$ (KL) versus $w_\ve^\ast=\sqrt{D_g\,\ve}$ (CoV); both are of order $\sqrt{\ve}$.

\section{Numerical Example: Stochastic Korteweg--de Vries (KdV) equation}\label{sec:kdv}

We study the stochastic KdV equation on the periodic domain $x\in[0,2\pi]$ for $t\in[0,1]$,
\begin{equation*}
\partial_t u + u \partial_x u - \nu \partial_{xx}u + \theta \partial_{xxx}u = \sqrt{\frac{\ve}{\pi}}\left( \sin (x) \dot{B}_1(t) + \cos (x)\dot{B}_2(t)\right),
\end{equation*}
with zero initial condition and $\nu = \theta = 4 \times 10^{-2}$. Here, $B_1$ and $B_2$ are independent standard Brownian motions, and $\ve > 0$ controls the noise strength. The stochastic forcing is applied only to the first spatial Fourier mode, while the nonlinear term generates the higher modes of the solution. This example was also used in \cite{schorlepp-tong-grafke-etal:2023} to study SORM approximations and their scalability. In that work, it was noted that the event set exhibits strong curvature along some directions, and that the SORM prefactor changes by several orders of magnitude depending on $z$, making this a challenging problem for IS.
After time discretization with $n_t$ steps, the normalized Brownian increments form a standard Gaussian vector, giving the random input
\begin{equation*}
\bs \eta_\ve = \sqrt{\ve}\bs \xi, \quad \bs\xi \sim \mathcal{N}(\bs 0, \text{Id}_N), \quad N = 2n_t.
\end{equation*}
Let $u(x,t;\bs\eta)$ be the numerical solution for input $\bs\eta$. We study the wave height at the origin at the final time, defining the parameter-to-event map and rare-event probability as
\begin{equation*}
F(\bs \eta) = u(0, 1; \bs \eta), \quad P_\ve(z) = \PP(F(\bs \eta_\ve) \geq z), \quad z = 8.39125.
\end{equation*}
We fix $z$ and vary $\ve$. Since the instanton $\bs\eta_z$ and the Hessian $H_z$ do not depend on $\ve$, they are computed once and shared by all samplers and noise strengths. All CoVs reported below are estimated from the sample mean and sample standard deviation of the $M = 10^4$ estimator contributions of a single run. We report the median of this estimate over $20$ independent runs. 

We solve the KdV equation using a Fourier discretization in space with $n_x$ points and an integrating-factor Heun scheme in time. Unless stated otherwise, we use $(n_x, n_t) = (256, 1000)$, corresponding to $N = 2000$.  We compute the instanton by minimizing the rate function using an augmented Lagrangian method, with gradients and Hessian-vector products obtained by differentiating the discretized forward map. We retain the 100 tangential Hessian eigenmodes with the largest absolute eigenvalues, found using an iterative eigenvalue solver as in \cite{schorlepp-tong-grafke-etal:2023}, to construct the quadratic boundary approximation and the \pecan\ covariance correction. The remaining tangential directions are sampled with unchanged Gaussian variances. For the bottom-branch density of \pecan\ IBIS, we use the base density $g_\beta(x) := \beta e^{-\beta x}$ with $\beta=0.76$. Following \cref{thm:scaled-left-proposal}, the bottom-branch probability is $w_\ve = \min\{1/2, \sqrt{\hat{D}\ve}\}$ with $\hat{D}$ an estimate of $D_g$ in \eqref{eq:scaled-left-density}; how $\beta$ and $\hat{D}$ are determined, and how sensitive the results are to $\beta$, is discussed at the end of this section. 

\begin{figure}[t]
    \centering
    \includegraphics[]{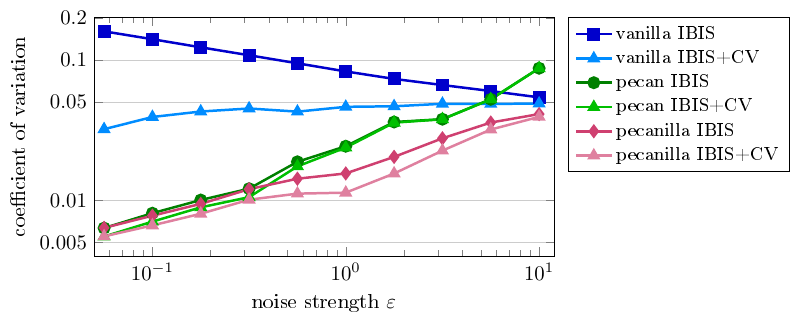}
    \vspace{-2ex}
\caption{KdV example:
Comparison of vanilla, \pecan, and \sundae\ IBIS, each with and without a control variate. 
The event-probability estimates are approximately $3.53 \times 10^{-4}$, $1.19 \times 10^{-7}$, $3.63\times 10^{-18}$, and $3.27 \times 10^{-154}$ for $\ve = 10$, $10^{0.5}$, $1$, and $0.1$, respectively. 
}
\label{fig:kdv-convergence}
\end{figure}

The comparison with subspace CE is presented in the motivating SDE example in \cref{fig:cov-comparison}. Here, we focus on the IBIS family to examine the effects of curvature correction, control variates, curvature damping, as well as robustness under discretization refinement.
\paragraph{Vanilla and \pecan\ samplers}
\Cref{fig:kdv-convergence} confirms the qualitative small-noise behavior predicted by the LDT-based analysis, similar to \cref{fig:cov-comparison}: the CoV of vanilla IBIS increases as $\ve$ decreases, its control-variate version remains roughly constant, and the CoV of \pecan\ IBIS decreases. Control variates yield only moderate improvement for \pecan\ IBIS. At the largest noise strength, vanilla IBIS outperforms \pecan\ IBIS, motivating the curvature damped variant introduced next. 

\paragraph{\Sundae\ IBIS}
As noted above and shown in \cite{schorlepp-tong-grafke-etal:2023}, this KdV example has a strongly shaped event boundary.
In particular, the projected Hessian $\lambda_z H_z$ has a large negative eigenvalue ($\approx -57.27$ at the baseline resolution). Since $G_z = \operatorname{Id} - \lambda_z H_z$ then has an eigenvalue of $\approx 58$, the \pecan\ tangential proposal $G_z^{-1/2} \bs u$ has a standard deviation $\approx 0.13$ in the corresponding direction. For small $\ve$, this matches the shape of the event set, but for larger $\ve$, the event set extends beyond the region where the quadratic approximation is accurate; the narrow proposal rarely reaches these parts and the few samples that do carry large importance weights. This explains the deterioration of \pecan\ IBIS for large $\ve$ in \cref{fig:kdv-convergence}.

To reduce this effect, we introduce \emph{\sundae} IBIS, which damps the curvature correction for large $\ve$ by replacing $H_z$ by $\gamma(\ve)H_z$ in both the tangential covariance and the proposal boundary, where 
\begin{equation*}
\gamma(\ve) = \frac{1}{1+ (\ve/\ve_c)^p}, \quad \ve_c = 1, \,\,p =2.
\end{equation*}
For $\ve \to \infty$, the sampler thus reduces to vanilla IBIS except for the piecewise-exponential proposal in the normal direction, while $\gamma(\ve) = 1 + O(\ve^2)$ leaves the small-noise results of \cref{sec:theory-cov,sec:theory-kl} unchanged. In \cref{fig:kdv-convergence}, \sundae\ IBIS matches \pecan\ IBIS for small $\ve$ and improves over both vanilla and \pecan\ IBIS for large $\ve$.

\begin{figure}[t]
    \centering
    \includegraphics[]{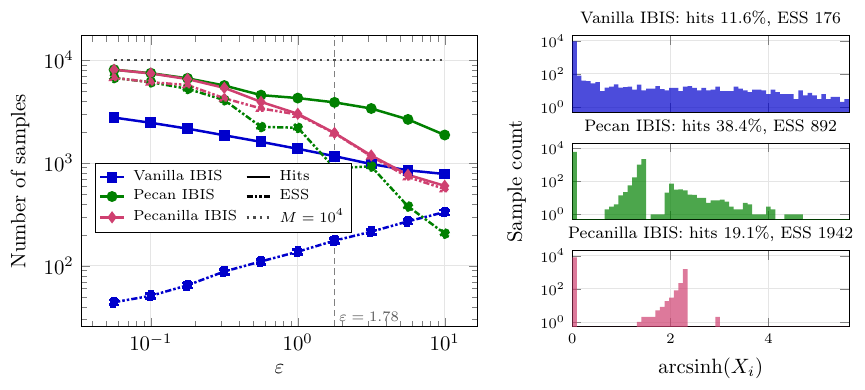}
    \vspace{-2ex}
\caption{
Left: event-hit counts (solid) and effective sample size (dashed) for vanilla, \pecan, and \sundae\ IBIS.
Right: histograms of $\operatorname{arcsinh}(X_i)$ from one run at $\varepsilon = 10^{0.25} \approx 1.78$ (vertical dashed line in the left panel). Here, $X_i$ is the nonnegative estimator contribution (including the event indicator) before applying the control variate. 
The zero bin contains samples outside the event set. Control variates do not change the proposals or hit counts.
}
\label{fig:kdv-hit-fraction}
\end{figure}

\paragraph{Effective sample size}
\Cref{fig:kdv-hit-fraction} compares event-hit counts and the effective sample size $\mathrm{ESS} = (\sum_i X_i)^2/\sum_i X_i^2$. For nonnegative contributions, the empirical CoV satisfies $\widehat{\Cov}^2 = \frac{M}{M-1}\left(\frac{M}{\mathrm{ESS}} - 1\right)$, so ESS and CoV provide equivalent measures of sampling efficiency, whereas hit counts do not: more hits need not imply a larger ESS. At $\ve \approx 1.78$, \sundae\ IBIS achieves more than twice the median ESS of \pecan\ IBIS ($1942$ versus $892$) with roughly half as many hits. The single-run histograms illustrate the greater variability of the \pecan\ importance weights, consistent with its lower sampling efficiency.

\paragraph{Control variate versions}
\Cref{tab:kdv-cv-correlation} lists the empirical correlation $\varrho$ between the estimator contribution and the control variable. For the optimal control-variate coefficient, the variance is reduced by the factor $1 - \varrho^2$. Since the mean is unchanged, the corresponding CoV reduction factor is $\sqrt{1 - \varrho^2}$. For vanilla IBIS at the medium resolution and $\ve = 0.1$, the reported correlation $\varrho = 0.96$ gives a factor of $0.28$, consistent with the ratio $3.93/14.07 \approx 0.28$ of the median CoVs with and without the control variate in \cref{tab:kdv-cov-resolutions}.
Further, its correlation is strongest at small $\ve$, as predicted by \cref{thm:population-cv-coefficient}: the control variable captures the leading $\ve^{-1/2}$ contribution to the variance. For \pecan\ IBIS, this contribution is already removed by the proposal, and the remaining variance arises in part from the boundary mismatch below the parabola, which the quadratic control variable does not capture; accordingly, the correlation is weaker and vanishes at $\ve=10$. \Sundae\ IBIS retains a stronger correlation at moderate and large $\ve$, since its damped proposal boundary leaves part of the curvature effect to the control variate, consistent with its larger gains in \cref{fig:kdv-convergence}. 

\begin{table}[t]
\centering
\caption{Correlation between each estimator and its control variate at resolution $(n_x,n_t)=(256,1000)$. 
}
\label{tab:kdv-cv-correlation}
\small
\begin{tabular}{r|c|c|c}
\toprule
$\log_{10}\varepsilon$ & Vanilla IBIS & \Pecan\ IBIS & \Sundae\ IBIS\\
\midrule
-1.00 & 0.957 & 0.504 & 0.518\\
-0.50 & 0.889 & 0.497  & 0.564\\
0.00 & 0.852 & 0.327  & 0.695\\
0.50 & 0.695  & 0.115 & 0.581\\
1.00 & 0.426  & 0.002 & 0.337\\
\bottomrule
\end{tabular}
\end{table}

\paragraph{Choice of the bottom-branch density and sensitivity}
\Cref{thm:scaled-left-proposal} suggests choosing $g$ to minimize $D_g$ and setting $w_\ve = \sqrt{D_g\,\ve}$. For $g_\beta(x) = \beta e^{-\beta x}$, 
$D_{g_\beta}$ can be estimated by Monte Carlo from samples of $(c_3(\bs u))_-$. To this end, we draw 128 independent standard Gaussian tangential samples $\bs u_i$ and, at a reference noise strength $\ve_0 = 0.025$, locate the event boundary $s_{\ve_0}^\ast(\bs u_i)$ in the curvature-adjusted coordinates by bisection along the normal direction. Then $t_i := \max\left\{{(b(\bs u_i) - s_{\ve_0}^\ast(\bs u_i))}/{\sqrt{\ve_0}}, 0  \right\}$ approximates $(c_3(\bs u_i))_-$ (cf.\ \cref{sec:mod-ibis-general}), and we minimize
\begin{equation*}
\widehat{D}(\beta) = \frac{1}{128} \sum_{i=1}^{128} \frac{e^{\beta t_i}-1}{\beta^2}, \quad \beta_\ast = \arg\min_{\beta > 0} \widehat{D}(\beta),
\end{equation*}
which yields $\beta_\ast \approx 0.76$ and $\widehat D(\beta_\ast)\approx 0.5$; the bottom-branch probability is $w_\ve = \min \{1/2,\allowbreak (\widehat{D}(\beta_\ast)\ve)^{1/2}\}$.  
Here, $\widehat{D}$ is an estimate of the typical size of $D_g$, i.e., of its value in the region that is actually explored by moderately many samples. 
To test sensitivity to $\beta$, we varied $\beta/\beta_\ast \in \{1/4, 1/2, 1, 2, 4\}$. At $\ve=0.1$, median ESS values were $4866$, $5827$, $6039$, $6762$, and $6470$, showing only moderate sensitivity. For different $\beta$, the CoV curves show similar qualitative dependence on $\ve$. Occasional gains over $\beta_\ast$ are expected because $\beta_\ast$ minimizes $\widehat{D}(\beta)$—a heuristic proxy for the leading small-noise variance coefficient—rather than the empirical CoV at finite $\ve$.

\paragraph{Scalability}
To test robustness with respect to resolution, \cref{tab:kdv-cov-resolutions} compares the CoV at $(n_x, n_t) = (128, 500)$, $(256, 1000)$, and $(512, 2000)$, i.e., $N = 1000$, $2000$, and $4000$, for $\ve = 0.1$, $1$, and $10^{0.5}$. The CoVs and the ranking of the methods are essentially unchanged from $N = 1000$ to $N = 4000$. This is expected: the constants in \cref{sec:theory-cov} are determinants and Gaussian expectations that converge as the discretization is refined~\cite{schorlepp-grafke:2025}, so we expect the CoV of all IBIS variants to have an $N$-independent limit; see \cref{sec:conclusion} for the caveats regarding the \pecan\ constants.

\begin{table}[t]
\centering
\caption{Resolution dependence of the estimated CoV for the KdV problem. Entries are $10^2$ times the median estimated CoV over $20$ repetitions
(thus $0.81$ denotes a CoV of $0.0081$).
Low, medium, and high resolutions correspond to $(n_x,n_t)=(128,500)$,
$(256,1000)$, and $(512,2000)$, respectively.
}
\label{tab:kdv-cov-resolutions}
\small
\setlength{\tabcolsep}{3.5pt}
\renewcommand{\arraystretch}{1.12}
\begin{tabular}{@{}l*{9}{r}@{}}
\toprule
& \multicolumn{3}{c}{$\varepsilon=0.1$} & \multicolumn{3}{c}{$\varepsilon=1$} & \multicolumn{3}{c}{$\varepsilon=10^{0.5}$} \\
\cmidrule(lr){2-4}\cmidrule(lr){5-7}\cmidrule(l){8-10}
Method & Low & Med & High & Low & Med & High & Low & Med & High \\
\midrule
Vanilla IBIS & 14.43 & 14.07 & 14.12 & 8.23 & 8.29 & 8.41 & 6.53 & 6.63 & 6.63 \\
Vanilla IBIS + CV & 4.34 & 3.93 & 3.46 & 4.66 & 4.64 & 4.34 & 4.61 & 4.88 & 4.81 \\
\Pecan\ IBIS & 0.93 & 0.81 & 0.80 & 1.83 & 2.43 & 1.88 & 4.21 & 3.79 & 3.53 \\
\Pecan\ IBIS + CV & 0.88 & 0.71 & 0.71 & 1.71 & 2.38 & 1.78 & 4.19 & 3.78 & 3.51 \\
\Sundae\ IBIS & 0.93 & 0.78 & 0.77 & 1.54 & 1.56 & 1.54 & 2.76 & 2.77 & 2.77 \\
\Sundae\ IBIS + CV & 0.88 & 0.66 & 0.67 & 1.12 & 1.13 & 1.12 & 2.25 & 2.26 & 2.26 \\
\bottomrule
\end{tabular}
\end{table}

\begin{figure}[t]
    \centering
    \includegraphics[width=0.83\textwidth]{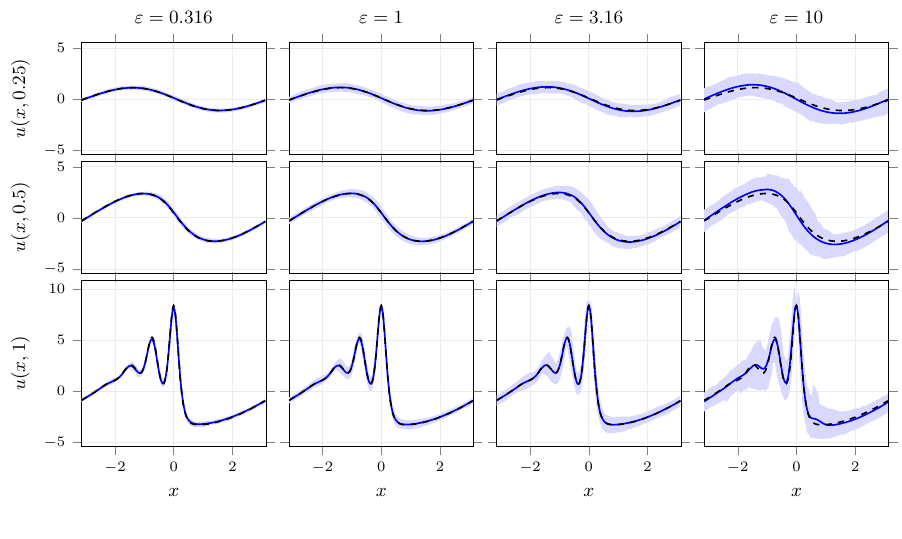}
    \vspace{-3.5ex}
    \caption{
        Spatial profiles of vanilla IBIS samples satisfying $|F(\bs \eta_i) - z| \leq 0.25$.
        Rows show KdV solution at $t = 0.25$, $0.5$, and $1$; columns show $\ve = 10^{-0.5}$, $1$, $10^{0.5}$, and $10$. Dashed black curves show the instanton, blue curves show importance-weighted sample means, and shaded regions show pointwise 5--95\% sample bands. The same selected trajectories are used at all three times within each column.
    }
    \label{fig:kdv-concentration}
\end{figure}

\paragraph{Conditional trajectories}
\Cref{fig:kdv-concentration} visualizes the conditional law of the trajectories given the rare event, which underlies the small-noise analysis: we select vanilla IBIS samples with $|F(\bs\eta_i)-z|\le0.25$ and show importance-weighted means and pointwise 5--95\% bands of the profiles $u(x,t;\bs\eta_i)$ at three times and four noise strengths. As $\ve$ decreases, the bands narrow and the profiles concentrate around the instanton, as assumed in \cref{sec:ldt-setup}. At larger $\ve$, the profiles deviate substantially from the instanton---at $t=1$ the band is wide around the secondary peak and to the right of the main peak, although the wave height at $x=0$ stays close to $z$---which is the regime in which the local quadratic approximation, and hence \pecan\ IBIS, loses accuracy (cf.\ \cref{fig:kdv-convergence}). Consistently, minimizing the rate function from different initial guesses identifies additional local minimizers, all of which have strictly larger rate function values than the instanton used here; these do not affect the small-noise asymptotics but contribute to the conditional law at large $\ve$. 

\section{Discussion and outlook}
\label{sec:conclusion}
Through a small-noise limit analysis, we showed that the CoV of vanilla IBIS diverges like $\ve^{-1/4}$, a control variate makes it bounded, and \pecan\ IBIS leads to a $\ve^{1/4}$ decay for truncated and localized versions of the respective estimators; our KL analysis reproduces two of these rates without second-moment assumptions, and for \pecan\ IBIS without localization. This yielded practical IS algorithms for rare-event estimation and inspired the fused \sundae\ IBIS method, which improves upon both samplers across a wide range of noise levels.

\begin{figure}
       \centering
        \includegraphics[]
            {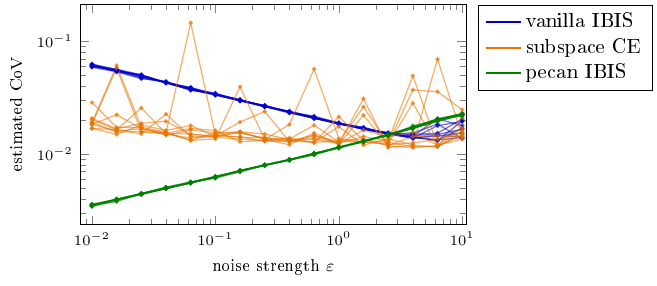}
       \vspace{-2ex}
        \caption{
    Run-to-run variability of the estimated CoV
 for the SDE \eqref{eq:2d-sde-intro}. Each curve represents one repetition. Vanilla and \pecan\ IBIS use $M=10^4$ samples per repetition. Subspace CE uses $M/2$ training and $M/2$ testing samples per run. CoVs are shown for 10 runs of each sampler. The $y$-axis differs from \cref{fig:cov-comparison} due to the smaller $M$ used here.
        }
        \label{fig:sde-cov-repetitions}
\end{figure}

We close with an observation on the subspace CE results of \cref{fig:cov-comparison}: reporting median CoVs hides variability across repetitions. To examine sampling variability separately from proposal fitting, \cref{fig:sde-cov-repetitions} shows individual repetitions: subspace CE exhibits occasional large empirical CoVs, while vanilla and \pecan\ IBIS are consistent. The following one-dimensional Gaussian tail example illustrates how a mismatch between the proposal and target tails can produce this behavior.
Consider estimating $P_\ve(z) = \mathbb{P}(X \geq z)$ for $X \sim \mathcal{N}(0, \ve)$ and fixed $z > 0$. The optimal IS distribution is the conditional law of $X$ given $X \geq z$. Gaussian CE minimizes the KL divergence from this conditional distribution by matching its mean and variance. With exact moments, the resulting proposal $q_\ve = \mathcal{N}(m_\ve, v_\ve)$ satisfies
$m_\ve - z \sim {\ve}/{z}$, $v_\ve \sim {\ve^2}/{z^2}$ as $\ve \downarrow 0$.
Thus, CE captures the narrow concentration near the event boundary. However, its Gaussian shape gives it a much thinner right tail than the conditional target: On the natural scale $y = z(x-z)/\ve$, the conditional target approaches an exponential distribution with rate one, whereas the CE proposal approaches $\mathcal{N}(1,1)$. 
For $v_\ve \leq \ve/2$, this is a more general case of the counterexample in \cref{sec:theory-kl}. The non-finite CoV, or $\chi^2$-divergence, is equivalent to occasional large importance weights. This is illustrated by small empirical CoVs in most repetitions coexisting with occasional large outliers, and explains why minimizing KL divergence does not ensure finite IS variance.
A KL-optimal Gaussian proposal is too thin-tailed for the exponential conditional law near the boundary, so the KL divergence stays finite while the variance does not. The exponential tilting of the IBIS family avoids this by construction, which suggests \pecan-type proposals as a desirable parametric family for CE or other adaptive IS methods.

Our results are formulated for fixed dimension $N$, while the KdV example suggests that the relevant quantities are essentially mesh-independent. The constants $\beta_{\rm CoV}$ and $\beta_{\rm KL}$ of vanilla IBIS in \cref{sec:theory-cov,sec:theory-kl} depend on the Hessian only through products over its eigenvalues with
factors $1+O(\kappa_i^2)$.
These constants remain bounded as $N\to\infty$ whenever $\sum_i\kappa_i^2<\infty$, i.e., for Hilbert--Schmidt (rather than trace-class) Hessians, in line with the Carleman--Fredholm determinant in \cite{schorlepp-grafke:2025}. The same holds for the limiting variance $1+\frac12\operatorname{tr}((\lambda_zH_zG_z^{-1})^2)
$ of the log weight in \cref{prop:vanilla-ibis-KL}. The constants of the control-variate and \pecan\ results ($C_{\CV}$, $m_1$, $D_g$, $\widetilde{D}_g$) additionally involve the cubic boundary correction $c_3$, i.e., third derivatives of $F$, whose behavior as $N\to\infty$ we have not analyzed. A rigorous extension to the infinite-dimensional setting would require, in addition, uniformity in $N$ of the $o(\sqrt{\ve})$ remainders and of the truncation to $\mathcal{B}_R$ (redefined in a suitably reweighted norm, e.g., to render \cref{fn:R-tube} dimension-independent), which we leave for future work.

On the practical side, the asymptotically optimal bottom-branch weights $w_\ve^\ast=\sqrt{D_g\ve}$ (CoV) and $w_\ve^\ast=m_1\sqrt{\ve}$ (KL) involve the generally unavailable cubic boundary expansion coefficient $c_3$. In \cref{sec:kdv} we thus calibrated the prefactor. Since, to leading order, the target places mass $m_1\sqrt{\ve}$ below the parabola, $m_1$ can be estimated directly from the fraction of event samples that fall below it, which suggests an adaptive choice of $w_\ve$; the same holds for the damping scale $\ve_c$ of \sundae\ IBIS. A modified control variate for \pecan\ IBIS, analogous to the one that removes the growth of vanilla IBIS, and the treatment of multiple or degenerate instantons, where the $\sqrt{\ve}$-expansions break down, are further natural directions. Finally, removing the tangential localization from the \pecan\ IBIS variance analysis requires heavy-tailed bottom-branch densities together with a global condition on the event set adapted to the \pecan\ proposal (see \cref{sec:mod-ibis-general}), which we also leave for future work.
More broadly, we hope that the divergence-based perspective complements moment-based error measures in the evaluation of LDT-based samplers.

\section*{Acknowledgments} The authors thank Claudia Strauch for helpful discussions. T.S.\ and G.S.\ thank Shanyin Tong for earlier related discussions in 2022.

\bibliographystyle{siamplain}
\bibliography{bib}

\appendix

\section{Derivations and proofs for CoV analysis}
\label{app:proofs-sec3}
This appendix provides the technical derivations and proofs of the results stated in \cref{sec:theory-cov}. Throughout, we use the notation and common assumptions introduced in \cref{sec:problem-setup}.

\begin{lemma}[Uniform separation from the instanton]
\label{lemma:uniform-separation}
Fix $\bs p \in \R^N$, and let $\bs \eta^\ast$ be the unique global minimizer of $\bs\eta \mapsto \frac{1}{2}\Vert \bs \eta -\bs p \Vert^2$ over the closed set $\{\bs\eta: F(\bs\eta) \geq z \}$. Then for every $\delta > 0$,
\begin{equation*}
\inf_{\bs \eta} \left\{\frac{1}{2}\Vert \bs\eta - \bs p \Vert^2: F(\bs\eta) \geq z, \Vert \bs\eta - \bs\eta^\ast \Vert \geq \delta  \right\} > \frac{1}{2}\Vert \bs\eta^\ast -  \bs p \Vert^2.
\end{equation*}
\end{lemma}
\begin{proof}
Suppose the infimum equals $\frac{1}{2}\Vert \bs\eta^\ast - \bs p \Vert^2$ (it cannot be smaller). Then there is a sequence $(\bs\eta_n)$ with $F(\bs\eta_n) \geq z$, $\Vert \bs\eta_n - \bs\eta^\ast \Vert \geq \delta$, and $\frac{1}{2}\Vert \bs\eta_n - \bs p \Vert^2 \to \frac{1}{2}\Vert \bs\eta^\ast - \bs p\Vert^2$. It is bounded, so a subsequence converges to some $\bar{\bs\eta}$ with $F(\bar{\bs\eta}) \geq z$ by continuity of $F$, $\Vert \bar{\bs\eta} - \bs\eta^\ast \Vert \geq \delta$, and $\frac{1}{2}\Vert \bar{\bs\eta} - \bs p \Vert^2 = \frac{1}{2}\Vert \bs\eta^\ast - \bs p\Vert^2$, i.e., a second global minimizer, contradicting uniqueness.
\end{proof}
We use this lemma, with $\bs p=\bs 0$ (\cref{assump:standard-ldt}(ii)) and with $\bs p=-\bs\eta_z$ (\cref{assump:variance}(i)), to show that the region outside any fixed neighborhood of the instanton contributes negligibly to the integrals below.
To further control the boundary approximation error, we will use the following consequence of \cref{assump:standard-ldt}. It is local in the original variables, i.e., it holds on a fixed ball $\Vert \bs\eta - \bs\eta_z \Vert \lesssim \delta_0$ around the instanton, which in the scaled tangential coordinate $\bs u$ corresponds to the growing region $\Vert \bs u \Vert \leq \delta_0/\sqrt{\ve}$.
\begin{lemma}[local third-order boundary expansion]
\label{lem:boundary-expansion}
Under \cref{assump:standard-ldt}, there exist $\delta_0, \ve_0, C > 0$ such that for all $\ve \in (0, \ve_0]$ and all $\bs u \in \bs\eta_z^\perp$ with $\Vert \bs u \Vert \leq \delta_0/\sqrt{\ve}$ the following holds: among the points~\eqref{eq:inst-local-coords} with $|\ve s| \leq \delta_0$, the event $\{F \geq z\}$ is exactly $\{s \geq s_\ve^\ast(\bs u)\}$ for a unique $s_\ve^\ast(\bs u)$, and
\begin{equation*}
\left| s_\ve^\ast(\bs u) - \tau(\bs u) - \sqrt{\ve}\, c_3(\bs u) \right| \leq C \ve \left(1 + \Vert \bs u \Vert\right)^4, \qquad \left| s_\ve^\ast(\bs u) - \tau(\bs u) \right| \leq C\left(\sqrt{\ve} + \delta_0\right) \left(1 + \Vert \bs u \Vert\right)^2,
\end{equation*}
where $c_3$ is the odd cubic polynomial given in~\eqref{eq:c3def}. In particular, $s_\ve^\ast(\bs u) = \tau(\bs u) + \sqrt{\ve}\,c_3(\bs u) + O(\ve)$ uniformly on every bounded tangential region $\mathcal{B}_R = \{\bs u: \Vert \bs u \Vert \leq R\}$. The same holds, with modified constants, in the curvature-adjusted coordinates of \cref{sec:mod-ibis-def}, with $b(\bs u)$ in place of $\tau(\bs u)$ and the correspondingly transformed $c_3$.
\end{lemma}
The proof relies on monotonicity of $F$ in the normal direction, with the Taylor remainder controlled by the Lipschitz third derivative:
\begin{proof}
Write $\rho: = \Vert \bs u \Vert$ and $f(s,\bs u):=(F(\bs\eta)-z)/\ve$ for $\bs\eta=\bs\eta_z+\ve s\bs e_z+\sqrt{\ve}\bs u$. For sufficiently small $\delta_0$, continuity of $\nabla F$ and the first-order optimality condition give $\partial_s f(s, \bs u) \geq r/(2\lambda_z)$ throughout $|\ve s| \leq \delta_0$ and $\sqrt{\ve} \rho \leq \delta_0$. Thus $f$ is strictly increasing in $s$ on this region. 

Set $s_0(\bs u) = \tau(\bs u) + \sqrt{\ve}c_3(\bs u)$ with $c_3$ given by~\eqref{eq:c3def}. Taylor expansion, using the locally Lipschitz third derivative and the cancellation defining $c_3$, yields
\[|f(s_0(\bs u), \bs u)| \leq C\ve(1+\rho)^4.\]
Since $\tau$ is quadratic and $c_3$ is cubic, an interval of radius $C \ve (1+\rho)^4$ around $s_0$, with $C$ sufficiently large, lies within the admissible normal range when $\delta_0$ and $\ve_0$ are sufficiently small. Monotonicity and the intermediate value theorem therefore give a unique zero satisfying
\[|s_\ve^\ast(\bs u) - s_0(\bs u) | \leq C \ve (1 + \rho)^4. \]
This proves the first estimate. The second follows from the cubic growth of $c_3$ and $\sqrt{\ve}(1+\rho) \leq \sqrt{\ve} + \delta_0$. Monotonicity also gives the stated event characterization. The curvature-adjusted version follows by the linear change of variables $s\mapsto s/r$, $\bs u\mapsto G_z^{-1/2}\bs u$.
\end{proof}

\subsection{Vanilla IBIS moments}
\label{app:vanilla-ibis-moments}

Let
\[
\varphi_\perp(\bs u):=(2\pi)^{-(N-1)/2}\exp\left(-\frac{1}{2}\Vert\bs u\Vert^2\right)\,, \quad \bs u \in \bs \eta_z^\perp\,,
\]
denote the standard Gaussian density on the $(N-1)$-dimensional subspace $\bs\eta_z^\perp \subset \mathbb{R}^N$, and recall 
\eqref{eq:inst-local-coords}, i.e., $\bs\eta=\bs\eta_z+\ve s\bs e_z+\sqrt{\ve}\bs u$, with $s\sim\mathcal{N}(0,1/\ve)$ and $\bs u\sim\mathcal{N}(\bs 0,\operatorname{Id}_{\bs\eta_z^\perp})$ independent.

\begin{proof}[Proof of \cref{prop:vanilla-ibis-moments}]
By the definition of $X_\ve$ in \cref{eq:ibis-def}, for $k\in\{1,2\}$,
\begin{equation*}
\E[X_\ve^k]
=\left(\frac{2\pi}{\ve}\right)^{\frac{k-1}{2}}\int_{\bs\eta_z^\perp}\int_{\mathbb{R}}\ind_{\{F(\bs\eta_z+\ve s\bs e_z+\sqrt{\ve}\bs u)\geq z\}}e^{-krs-\ve s^2/2}\,\mathrm{d}s\,\varphi_\perp(\bs u)\,\mathrm{d}\bs u.
\end{equation*}
The Taylor expansion in \cref{eq:quad-boundary-def} shows that, as $\ve\downarrow0$, the event condition becomes
$
s\geq\tau(\bs u)
$
with
$\tau(\bs u)=-\frac{\lambda_z}{2r}\langle\bs u,H_z\bs u\rangle$.
The precise version of this statement, under \cref{assump:standard-ldt}, is given by \cref{lem:boundary-expansion}.
In the original variables, $\E[X_\ve]\propto\int_{\{F\geq z\}}e^{-\|\bs\eta\|^2/(2\ve)}\,\mathrm{d}\bs\eta$ and $\E[X_\ve^2]\propto\int_{\{F\geq z\}}e^{-\|\bs\eta+\bs\eta_z\|^2/(2\ve)}\,\mathrm{d}\bs\eta$, whose exponents are the rate functions \eqref{eq:rate-func-def} and \eqref{eq:variance-rate-function}, respectively. By \cref{lemma:uniform-separation} with $\bs p=\bs 0$ (\cref{assump:standard-ldt}(ii)) and $\bs p=-\bs\eta_z$ (\cref{assump:variance}(i)), both integrals concentrate at $\bs\eta_z$.
The region outside $\{|\ve s| \leq \delta_0,\ \Vert \bs u \Vert \leq \delta_0/\sqrt{\ve}\}$, which contains the fixed ball $\Vert \bs\eta - \bs\eta_z \Vert \leq \delta_0$, contributes only an exponentially small relative amount. Inside this region, \cref{lem:boundary-expansion} gives $|s_\ve^\ast(\bs u) - \tau(\bs u)| \leq C(\sqrt{\ve} + \delta_0)(1 + \Vert \bs u \Vert)^2$, so that the inner $s$-integral is bounded by $(kr)^{-1}\exp\big(-kr\tau(\bs u) + C'(\sqrt{\ve}+\delta_0)(1+\Vert\bs u\Vert)^2\big)$. Since $\operatorname{Id}_{\bs\eta_z^\perp} - k\lambda_z H_z \succ 0$ for $k = 1$ (\cref{assump:standard-ldt}(iv)) and $k = 2$ (\cref{assump:variance}(ii)), this is integrable against $\varphi_\perp$ for $\delta_0$ and $\ve$ small enough, and dominated convergence applies.
Thus, the leading order follows by Gaussian integration near the instanton.
We obtain
\begin{align*}
\E[X_\ve^k]
&\overset{\ve \downarrow 0}{\sim}\left(\frac{2\pi}{\ve}\right)^{\frac{k-1}{2}}\int_{\bs\eta_z^\perp}\left[\int_{\tau(\bs u)}^\infty e^{-krs}\,\mathrm{d}s\right]\varphi_\perp(\bs u)\,\mathrm{d}\bs u\\
&=\left(\frac{2\pi}{\ve}\right)^{\frac{k-1}{2}}\frac{1}{kr}\int_{\bs\eta_z^\perp}\exp\left(\frac{k\lambda_z}{2}\langle\bs u,H_z\bs u\rangle\right)\varphi_\perp(\bs u)\,\mathrm{d}\bs u =\left(\frac{2\pi}{\ve}\right)^{\frac{k-1}{2}}\frac{1}{k}\,C_{\rm SORM}(z,k\lambda_z),
\end{align*}
where the Gaussian integral equals $\det(\operatorname{Id}_{\bs\eta_z^\perp}-k\lambda_zH_z)^{-1/2}$, which is finite since the matrix is positive definite due to \cref{assump:standard-ldt}(iv) for $k=1$, and \cref{assump:variance}(ii) for $k=2$. The last step uses $r=\sqrt{2I(z)}$ and \cref{eq:sorm-prefac-def}. For $k=1$ this is part~(i). For $k=2$,
since $\E[X_\ve]^2=O(1)$ by part~(i), the claimed asymptotics hold for $\Var(X_\ve)=\E[X_\ve^2]-\E[X_\ve]^2$, which is part~(ii).
\end{proof}

\subsection{Vanilla IBIS control-variable identities and truncation}
\label{app:control-variable-identities}
This subsection establishes the exact mean and infinite variance of the control variable. We then give the probability bound relating the original and truncated samples and estimate the change in their means.
\begin{lemma}[Control variable mean]
\label{lem:control-variable-mean}
For every $\ve > 0$, 
$\E[Y_\ve] = C_{\rm SORM}(z, \lambda_z)$.
\end{lemma}
\begin{proof}
Conditioning on $\bs u$ and using the definition of $Y_\ve$ in \cref{eq:control-variable-def}, we obtain
\begin{align*}
\E[Y_\ve\mid\bs u]
&=\sqrt{\frac{2\pi}{\ve}}\int_{\tau(\bs u)}^\infty e^{-rs+\ve s^2/2}\sqrt{\frac{\ve}{2\pi}}e^{-\ve s^2/2}\,\mathrm{d}s =\int_{\tau(\bs u)}^\infty e^{-rs}\,\mathrm{d}s
=\frac{1}{r}e^{-r\tau(\bs u)}.
\end{align*}
Therefore,
\begin{align*}
\E[Y_\ve]
&=\frac{1}{r}\E\left[e^{-r\tau(\bs u)}\right]
=\frac{1}{r}\E\left[\exp\left(\frac{\lambda_z}{2}\langle\bs u,H_z\bs u\rangle\right)\right]=\frac{1}{r}\det\left(\operatorname{Id}_{\bs\eta_z^\perp}-\lambda_zH_z\right)^{-1/2}
=C_{\rm SORM}(z,\lambda_z),
\end{align*}
where the last step uses $r=\sqrt{2I(z)}$ and \cref{eq:sorm-prefac-def}. This calculation is exact for every $\ve>0$.
\end{proof}

\begin{proposition}[Control variable population variance]
\label{prop:control-variable-var}
For every fixed $\ve > 0$,
$\E[Y_\ve^2] = \infty$,
and consequently,
$\Var(Y_\ve) = \infty$.
\end{proposition}
\begin{proof}
Squaring $Y_\ve$ gives
$
Y_\ve^2 = (2\pi/\ve)\,\ind_{\{s\geq\tau(\bs u)\}}\exp\left(-2rs+\ve s^2\right)
$.
Conditioning on $\bs u$, we obtain
\[
\E[Y_\ve^2\mid\bs u]=\sqrt{\frac{2\pi}{\ve}}\int_{\tau(\bs u)}^\infty\exp\left(-2rs+\frac{\ve s^2}{2}\right)\,\mathrm{d}s.
\]
For every fixed $\ve>0$, the positive quadratic term $\ve s^2/2$ dominates the linear term $-2rs$ as $s\to\infty$. Consequently, the integrand grows without bound and the integral diverges. Thus,
$
\E[Y_\ve^2\mid\bs u]=\infty
$
for every $\bs u$, and hence $\E[Y_\ve^2]=\infty$. Since $\E[Y_\ve]$ is finite by \cref{lem:control-variable-mean}, it follows that $\Var(Y_\ve)=\infty$.
We further note that even after imposing any finite upper cutoff $B$ for $s$, as soon as $\kappa_i > 0$ for at least one of the eigenvalues of $\lambda_z H_z$, we also have $\E\left[Y_\ve^2 \ind_{s \leq B} \right] = \infty$ at any fixed $\ve  > 0$ because of the integral's behavior at large negative $s$.
\end{proof}

The exponent in the preceding second-moment integral is
$
\psi_\ve(s):=-2rs+{\ve s^2}/{2}.
$
Since $\psi_\ve'(s)=-2r+\ve s$, the exponent begins increasing at $s = 2r/\ve$. Under the proposal, $s$ has standard deviation $1/\sqrt{\ve}$, so this point lies $2r/\sqrt{\ve}$ standard deviations above the mean.

\begin{figure}
    \centering
    \includegraphics[width=.75\textwidth]{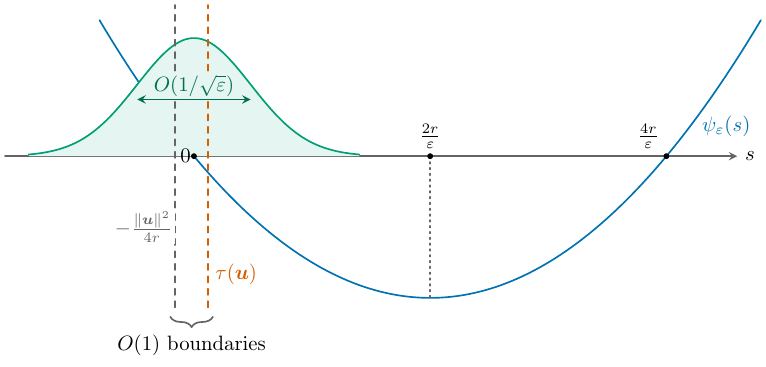}

    \caption{Illustration of the variance computation for the control variable $Y_\ve$ given $\bs u$. The exponent in the variance integral is $\psi_{\ve}(s) :=  -2rs + \ve s^2/2$ (blue line), and we evaluate $\int_{\tau(\bs u)}^\infty \exp(\psi_\ve(s)) \mathrm{d}s$. Since $s \sim {\cal N}(0,1/\ve)$ (green curve), and typical samples of $s$ hence lie at $O(1 / \sqrt{\ve})$, one never sees the actual blow-up of the variance, due to the integrand at $s = O(1/\ve)$, at moderate sample sizes with high probability. Under \cref{assump:variance}, $\tau(\bs u) \geq -\| \bs u\|^2 / (4r)$ for all $\bs u$, so that large negative $s$ are admissible only for large $\Vert \bs u\Vert$.}
    \label{fig:cv-sketch}
\end{figure}
Recall the cutoff $b_\ve = a r/\ve$, with $a$ as in \cref{sec:vanilla-ibis-cv-general}, and define $\mathcal{A}_{\ve, M} = \left\{\max_{1 \leq m \leq M} |s^{(m)}| \leq b_\ve \right\}$, which is the event that all sampled normal coordinates lie within the cutoff.

\begin{lemma}[Probability of extreme sample]
\label{lem:extreme-sample}
The probability of the complementary event satisfies
\begin{equation*}
\PP(\mathcal{A}_{\ve, M}^c) \leq \frac{2 M \sqrt{\ve}}{a r \sqrt{2\pi}}\exp\left( - \frac{a^2 r^2}{2\ve}\right).
\end{equation*}
Therefore, if
$
M(\ve) \sqrt{\ve}\exp\left( - a^2 r^2 / (2\ve)\right) \longrightarrow 0,
$
then $\mathbb{P}(\mathcal{A}_{\ve, M}) \longrightarrow 1$.
\end{lemma}
\begin{proof}
Let $s^{(1)},\ldots,s^{(M)}$ be the independent samples used by the estimator. Since $\sqrt{\ve}s^{(m)}\sim\mathcal{N}(0,1)$, the Gaussian tail bound gives
\begin{align*}
\PP\left(|s^{(m)}|>b_\ve\right)
&=\PP\left(\sqrt{\ve}|s^{(m)}|>\frac{a r}{\sqrt{\ve}}\right)\leq\frac{2\sqrt{\ve}}{a r\sqrt{2\pi}}\exp\left(-\frac{a^2r^2}{2\ve}\right).
\end{align*}
Using the union bound over the $M$ samples, we obtain
\begin{align*}
\PP(\mathcal{A}_{\ve,M}^c)
&=\PP\left(\max_{1\leq m\leq M} |s^{(m)}| >b_\ve\right)\leq\sum_{m=1}^M\PP\left(|s^{(m)}|>b_\ve\right)\leq\frac{2 M\sqrt{\ve}}{a r\sqrt{2\pi}}\exp\left(-\frac{a^2r^2}{2\ve}\right).
\end{align*}
\end{proof}
Recall the truncated variables $X_{\ve, a}$ and $Y_{\ve, a}$ from \cref{sec:vanilla-ibis-cv-general}. On $\mathcal{A}_{\ve, M}$, the truncated and original samples coincide, $X_{\ve,a}^{(m)}=X_\ve^{(m)}$ and $Y_{\ve,a}^{(m)}=Y_\ve^{(m)}$ for $m=1,\ldots,M$. Moreover, $\E[Y_\ve\ind_{\{s>b_\ve\}}\mid\bs u]=\int_{\max\{\tau(\bs u),b_\ve\}}^\infty e^{-rs}\,\mathrm{d}s\le e^{-rb_\ve}/r$, and the lower cutoff gives $\E[Y_\ve\ind_{\{s<-b_\ve\}}]\le r^{-1}\E[e^{-r\tau(\bs u)}\ind_{\{\tau(\bs u)<-b_\ve\}}]\le r^{-1}e^{-\theta a r^2/\ve}\,\E[e^{-(1+\theta)r\tau(\bs u)}]$ for any $\theta>0$ (Chernoff), where the last expectation $\det(\operatorname{Id}_{\bs\eta_z^\perp}-(1+\theta)\lambda_zH_z)^{-1/2}$ is finite for small $\theta>0$ by \cref{assump:standard-ldt}(iv).
Taken together, we thus have $0\le\E[Y_\ve]-\E[Y_{\ve,a}]\le r^{-1}e^{-ar^2/\ve} +O(e^{-\theta a r^2/\ve})$.
For $X_\ve$, \cref{lemma:uniform-separation} gives the same exponential smallness, since the discarded region satisfies $\Vert \bs \eta - \bs \eta_z \Vert \geq ar$. Thus both means change by exponentially small amounts. Replacing $\E[Y_{\ve, a}]$ by the known mean $\E[Y_\ve]$ leaves the variance unchanged and changes the estimator's mean only by an exponentially small amount, so the asymptotic conclusions of \cref{thm:cv-variance-reduction} remain valid.

\subsection{Vanilla IBIS control-variate asymptotics}
\label{app:control-variate-asymptotics}
We first derive the cubic correction to the quadratic event boundary, then establish the moment expansions behind \cref{thm:population-cv-coefficient,thm:cv-variance-reduction}, and show that the control-variate coefficient fitted from samples converges. Throughout, $X_{\ve,a}$ and $Y_{\ve,a}$ are the truncated variables of \cref{sec:vanilla-ibis-cv-general}, with the normal coordinate restricted to $|s|\le b_\ve=ar/\ve$, with $a$ as in \cref{sec:vanilla-ibis-cv-general}. The moment calculations are carried out directly under the full Gaussian law of $\bs u$, using \cref{lem:boundary-expansion} near the instanton and \cref{lemma:uniform-separation} away from it.

Let $A_z:=\nabla^2F(\bs\eta_z)$ and $\mathcal{T}_z:=\nabla^3F(\bs\eta_z)$. With $\bs\eta=\bs\eta_z+\ve s\bs e_z+\sqrt{\ve}\bs u$, a Taylor expansion around $\bs\eta_z$ sorted in powers of $\sqrt{\ve}$ gives
\begin{align*}
F(\bs\eta)-z
&=\ve\left[\frac{r}{\lambda_z}s+\frac{1}{2}\langle\bs u,H_z\bs u\rangle\right]+\ve^{3/2}\left[s\langle\bs e_z,A_z\bs u\rangle+\frac{1}{6}\mathcal{T}_z[\bs u,\bs u,\bs u]\right]+O(\ve^2).
\end{align*}
The bracket of order $\ve$ vanishes at $s=\tau(\bs u)$. Inserting $s=s_\ve^\ast(\bs u)=\tau(\bs u)+\sqrt{\ve}\,c_3(\bs u)+O(\ve)$ from \cref{lem:boundary-expansion} into $F(\bs\eta)=z$ and collecting the terms of order $\ve^{3/2}$ yields
\begin{align}
c_3(\bs u)=-\frac{\lambda_z}{r}\left[\tau(\bs u)\langle\bs e_z,A_z\bs u\rangle+\frac{1}{6}\mathcal{T}_z[\bs u,\bs u,\bs u]\right].
\label{eq:c3def}
\end{align}
Since $\tau$ is even in $\bs u$, while $\langle\bs e_z,A_z\bs u\rangle$ and $\mathcal{T}_z[\bs u,\bs u,\bs u]$ are odd, $c_3$ is odd. The weight $q(\bs u):=e^{-2r\tau(\bs u)}=e^{\lambda_z\langle\bs u,H_z\bs u\rangle}$, defined before \cref{thm:cv-variance-reduction}, is even.

\begin{proposition}[Moments of the control variate]
\label{prop:cv-moments}
Suppose \cref{assump:variance} holds, let $a$ be as in \cref{sec:vanilla-ibis-cv-general}, and set
\[
\beta := C_{\rm SORM}(z,\lambda_z),\quad
\tilde{C} := \sqrt{\frac{\pi}{2}}\,C_{\rm SORM}(z,2\lambda_z),\quad
m_3 := \E\big[q(\bs u)(c_3(\bs u))_+\big] = \tfrac12\E\big[q(\bs u)|c_3(\bs u)|\big].
\]
Then, as $\ve \downarrow 0$, for some $c>0$,
\begin{align*}
\E[X_{\ve,a}] &= \beta + O(\ve), \qquad \E[Y_{\ve,a}] = \beta + O(e^{-c/\ve}),\\
\Var(X_{\ve,a}) &= \Var(Y_{\ve,a}) + O(\sqrt{\ve}) = \tilde{C}\,\ve^{-1/2} - \beta^2 + O(\sqrt{\ve}),\\
\cov(X_{\ve,a},Y_{\ve,a}) &= \tilde{C}\,\ve^{-1/2} - \beta^2 - \sqrt{2\pi}\,m_3 + O(\sqrt{\ve}).
\end{align*}
\end{proposition}
\begin{proof}
Let $\delta_0,\ve_0,C$ be as in \cref{lem:boundary-expansion}, with $\delta_0\leq ar$ (decreasing $\delta_0$ if necessary), and let $\Omega_\ve:=\{|\ve s|\leq\delta_0,\ \Vert\bs u\Vert\leq\delta_0/\sqrt{\ve}\}$. We denote by $\Psi$ a generic function of the form
\[
\Psi(\bs u)=C'\left(1+\Vert\bs u\Vert\right)^8\left(1+e^{-(2+a/2)r\tau(\bs u)}\right)e^{C'(\sqrt{\ve_0}+\delta_0)(1+\Vert\bs u\Vert)^2},
\]
with $C'$ independent of $\ve$ and $\bs u$. Since $e^{-(2+a/2)r\tau(\bs u)}=e^{(1+a/4)\lambda_z\langle\bs u,H_z\bs u\rangle}$ and $\operatorname{Id}_{\bs\eta_z^\perp}-(2+a/2)\lambda_zH_z\succ0$ by the choice of $a$, $\Psi$ is integrable against $\varphi_\perp$ for $\delta_0$ and $\ve_0$ small enough, with Gaussian tails, i.e., $\E[\Psi(\bs u)\ind_{\{\Vert\bs u\Vert>\delta_0/\sqrt{\ve}\}}]=O(e^{-c/\ve})$.

\textit{Outside $\Omega_\ve$.} Since $\Omega_\ve$ contains the ball $\Vert\bs\eta-\bs\eta_z\Vert\leq\delta_0$, \cref{lemma:uniform-separation} with $\bs p=\bs 0$ and $\bs p=-\bs\eta_z$ shows $\E[X_{\ve,a}^k\ind_{\Omega_\ve^c}]\leq\E[X_\ve^k\ind_{\Omega_\ve^c}]=O(e^{-c/\ve})$ for $k=1,2$, as in the proof of \cref{prop:vanilla-ibis-moments}. For $Y_{\ve,a}$, we use $\ve s^2/2\leq ar|s|/2$ for $|s|\leq b_\ve$: the exponent $-2rs+\ve s^2/2$ in $\E[Y_{\ve,a}^2\mid\bs u]$ is at most $-(2-a/2)rs$ for $s\geq0$, and at most $(2+a/2)r|s|\leq-(2+a/2)r\tau(\bs u)$ for $\tau(\bs u)\leq s<0$. Hence $\E[Y_{\ve,a}^2\mid\bs u]\leq\ve^{-1/2}\Psi(\bs u)$, and the part with $s>\delta_0/\ve$ is $O(e^{-c/\ve})$. Moreover, for $\Vert\bs u\Vert\leq\delta_0/\sqrt{\ve}$, we have $|\tau(\bs u)|\leq C'\delta_0^2/\ve<\delta_0/\ve$ for $\delta_0$ small, so that $Y_{\ve,a}$ vanishes for $s<-\delta_0/\ve$ there. Thus $\E[Y_{\ve,a}^2\ind_{\Omega_\ve^c}]=O(e^{-c/\ve})$, and the remaining first and mixed moments on $\Omega_\ve^c$ are exponentially small by Cauchy--Schwarz.

\textit{Inside $\Omega_\ve$.} Fix $\Vert\bs u\Vert\leq\delta_0/\sqrt{\ve}$. By \cref{lem:boundary-expansion}, the event within $\Omega_\ve$ is $\{s\geq s_\ve^\ast(\bs u)\}$, and, since $\delta_0\leq ar$, the cutoff $b_\ve$ is inactive. Conditioning on $\bs u$, the second moments restricted to $\Omega_\ve$ are thus
\begin{equation*}
\sqrt{\tfrac{2\pi}{\ve}}\int_{\ell}^{\delta_0/\ve}e^{-2rs+\sigma\ve s^2/2}\,\mathrm{d}s,\qquad (\sigma,\ell)=\begin{cases}(-1,\,s_\ve^\ast(\bs u)) & \text{for } X_{\ve,a}^2,\\ (+1,\,\tau(\bs u)) & \text{for } Y_{\ve,a}^2,\\ (0,\,\max\{s_\ve^\ast(\bs u),\tau(\bs u)\}) & \text{for } X_{\ve,a}Y_{\ve,a},\end{cases}
\end{equation*}
and by \cref{lem:boundary-expansion}, $\ell=\tau(\bs u)+\sqrt{\ve}\,c(\bs u)+\varrho_\ve(\bs u)$ with $c=c_3$, $c=0$, and $c=(c_3)_+$, respectively, where $|\varrho_\ve(\bs u)|\leq C\ve(1+\Vert\bs u\Vert)^4$ and $|\ell-\tau(\bs u)|\leq C(\sqrt{\ve}+\delta_0)(1+\Vert\bs u\Vert)^2$. 

Replacing the quadratic exponential factor by 1 changes the inner integral by $O(\ve \Psi(\bs u))$, while extending its upper limit to infinity adds an exponentially small error. Expanding the resulting exponential integral at its lower limit, using the bounds from \cref{lem:boundary-expansion}, therefore gives
\[
\sqrt{\tfrac{2\pi}{\ve}}\int_{\ell}^{\delta_0/\ve}e^{-2rs+\sigma\ve s^2/2}\,\mathrm{d}s=\sqrt{\tfrac{2\pi}{\ve}}\,\frac{q(\bs u)}{2r}-\sqrt{2\pi}\,q(\bs u)\,c(\bs u)+O\big(\sqrt{\ve}\,\Psi(\bs u)\big).
\]
In the same way, $\E[X_{\ve,a}\ind_{\Omega_\ve}\mid\bs u]=\int_{s_\ve^\ast(\bs u)}^{\delta_0/\ve}e^{-rs-\ve s^2/2}\,\mathrm{d}s=\frac1r e^{-r\tau(\bs u)}\big(1-r\sqrt{\ve}\,c_3(\bs u)\big)+O(\ve\Psi(\bs u))$.

\textit{Averaging over $\bs u$.} We integrate these expansions against $\varphi_\perp$ over $\Vert\bs u\Vert\leq\delta_0/\sqrt{\ve}$, and extend the integrals to all of $\bs\eta_z^\perp$ at the cost $O(e^{-c/\ve})$ by the Gaussian tails of $\Psi$. Since $q$ and $e^{-r\tau}$ are even and $c_3$ is odd, the terms with $c=c_3$ vanish, while $c=(c_3)_+$ gives $-\sqrt{2\pi}\,m_3$. With $\E[e^{-r\tau(\bs u)}]=r\beta$ and $\E[q(\bs u)]=\det(\operatorname{Id}_{\bs\eta_z^\perp}-2\lambda_zH_z)^{-1/2}=rC_{\rm SORM}(z,2\lambda_z)$, finite by \cref{assump:variance}(ii), we obtain $\E[X_{\ve,a}]=\beta+O(\ve)$, $\E[X_{\ve,a}^2]=\E[Y_{\ve,a}^2]+O(\sqrt{\ve})=\tilde{C}\ve^{-1/2}+O(\sqrt{\ve})$, and $\E[X_{\ve,a}Y_{\ve,a}]=\tilde{C}\ve^{-1/2}-\sqrt{2\pi}\,m_3+O(\sqrt{\ve})$. Finally, $\E[Y_{\ve,a}]=\beta+O(e^{-c/\ve})$ by \cref{lem:control-variable-mean} and Appendix~\ref{app:control-variable-identities}, and subtracting the products of the means gives the claim.
\end{proof}

\begin{proposition}[Asymptotics of population control-variate coefficient]
\label{thm:population-cv-coefficient}
Suppose \cref{assump:variance} holds, and let $a$ be as in \cref{sec:vanilla-ibis-cv-general}. Then
\[
\Var(X_{\ve,a}),\ \Var(Y_{\ve,a}),\ \cov(X_{\ve,a},Y_{\ve,a}) = \tilde{C} \ve^{-1/2}+O(1), \qquad \tilde{C}:=\sqrt{\pi/2}\,C_{\rm SORM}(z,2\lambda_z).
\]
In particular, the optimal population control-variate coefficient~\eqref{eq:cv-coeff-opt} is $c_{\ve,a}^\ast = -1+O(\sqrt{\ve})$.
\end{proposition}
\begin{proof}
This follows directly from \cref{prop:cv-moments}; the coefficient follows from $c^\ast_{\ve,a}=-\cov(X_{\ve,a},Y_{\ve,a})/\Var(Y_{\ve,a})$.
\end{proof}

\begin{proposition}[Convergence of empirical control-variate coefficient]
\label{thm:empirical-cv-coefficient}
Suppose \cref{assump:standard-ldt} holds, and assume the dominating point condition~\eqref{eq:dominating-point}. Fix $0 < a < 2$.
Let $\hat{c}_{\ve, M}$ be the control-variate coefficient obtained from the empirical covariance of $X_\ve$ and $Y_\ve$ and the empirical variance of $Y_\ve$ over $M$ iid samples. Suppose that $M = M(\ve)$ satisfies
\begin{equation*}
   M(\ve) \ve \to \infty \quad \text{and} \quad  M(\ve) \sqrt{\ve} \exp\left( -\frac{a^2 r^2}{2\ve}\right) \to 0.
\end{equation*}
Then
$\hat{c}_{\ve, M(\ve)} \rightarrow -1$
in probability as $\ve\downarrow 0$.
\end{proposition}
The first condition provides enough samples for the empirical moments to concentrate (i.e., the empirical truncated coefficient $\hat{c}_{\ve, a}$ approaches the population coefficient $c_{\ve, a}^\ast$ in probability), and the second ensures that the probability of encountering the increasing region for the exponent vanishes. 
\begin{proof}[Proof sketch]
Let $U_{\ve,a}:=X_{\ve,a}Y_{\ve,a}$ and $V_{\ve,a}:=Y_{\ve,a}^2$. Under the dominating point condition, $\tau(\bs u)\ge0$, so whenever these variables are nonzero, $0\le\tau(\bs u)\le s\le b_\ve=ar/\ve$; since $0<a<2$, the exponents $-2rs+\ve s^2=s(\ve s-2r)$ and $-2rs+\ve s^2/2$ are then nonpositive, so $0\le U_{\ve,a},V_{\ve,a}\le2\pi/\ve$. By \cref{thm:population-cv-coefficient}, $\E[U_{\ve,a}]$ and $\E[V_{\ve,a}]$ equal $\tilde{C}\ve^{-1/2}+O(1)$, and Hoeffding's inequality for the empirical averages $\overline{U}_{\ve,a,M}$, $\overline{V}_{\ve,a,M}$ gives, for any $\delta>0$,
\[
\PP\left(\left|\frac{\overline{W}_{\ve,a,M}}{\E[W_{\ve,a}]}-1\right|>\delta\right)\le2\exp\left(-\frac{\delta^2\tilde{C}^2}{2\pi^2}\,M\ve\,[1+o(1)]\right),\qquad W\in\{U,V\},
\]
which tends to zero since $M(\ve)\ve\to\infty$. 

The empirical means $\overline{X}_{\ve,a,M}$, $\overline{Y}_{\ve,a,M}$ converge in probability to their bounded population means, because $\Var(X_{\ve,a}),\Var(Y_{\ve,a})=O(\ve^{-1/2})$ implies $\Var(\overline{X}_{\ve,a,M}),\Var(\overline{Y}_{\ve,a,M})=O(1/(M\sqrt{\ve}))\to0$. Hence $\widehat{\cov}_M(X_{\ve,a},Y_{\ve,a})=\overline{U}_{\ve,a,M}-\overline{X}_{\ve,a,M}\overline{Y}_{\ve,a,M}$ and $\widehat{\Var}_M(Y_{\ve,a})=\overline{V}_{\ve,a,M}-\overline{Y}_{\ve,a,M}^{\,2}$ are both $\tilde{C}\ve^{-1/2}[1+o_{\PP}(1)]$, and their ratio $\widehat{c}_{\ve,a,M}\to-1$ in probability. 

Finally, on $\mathcal{A}_{\ve,M}$ the truncated and original samples coincide, so $\widehat{c}_{\ve,M}=\widehat{c}_{\ve,a,M}$, and $\PP(\mathcal{A}_{\ve,M})\to1$ by \cref{lem:extreme-sample} and the second condition on $M(\ve)$.
\end{proof}

\begin{proof}[Proof of \cref{thm:cv-variance-reduction}]
By \cref{prop:cv-moments}, with $A:=\tilde{C}\ve^{-1/2}-\beta^2$ and $d:=\sqrt{2\pi}\,m_3$,
\[
\frac{\Var(X_{\ve,a}^{\CV})}{\Var(X_{\ve,a})}
=1-\frac{\cov(X_{\ve,a},Y_{\ve,a})^2}{\Var(X_{\ve,a})\Var(Y_{\ve,a})}
=1-\frac{(A-d+O(\sqrt\ve))^2}{(A+O(\sqrt\ve))^2}
=\frac{2d}{A}+O(\ve)
=\frac{2\sqrt{2\pi}\,m_3}{\tilde{C}}\sqrt{\ve}+O(\ve),
\]
and $2\sqrt{2\pi}\,m_3/\tilde{C}=2r\,\E[q(\bs u)|c_3(\bs u)|]/\E[q(\bs u)]=C_{\CV}$. For the CoV, $\Var(X^{\CV}_{\ve,a})\to C_{\CV}\tilde{C}$ and $\E[X^{\CV}_{\ve,a}]=\E[X_{\ve,a}]\to\beta=C_{\rm SORM}(z,\lambda_z)$, which gives the stated limit.
\end{proof}

\subsection{\Pecan\ IBIS moments}
\label{app:modified-ibis-moments}

Recall from \cref{sec:mod-ibis-def} that $s_\ve^\ast(\bs u)$ is the exact event boundary, $b(\bs u)$ is its quadratic approximation, and
$
\delta_\ve(\bs u):=s_\ve^\ast(\bs u)-b(\bs u)
$
the difference between the two at finite $\ve$.
We work on the bounded region $\mathcal{B}_R=\{\bs u:\Vert\bs u\Vert\leq R\}$ and use, by \cref{lem:boundary-expansion}, uniformly on $\mathcal{B}_R$ that
$
\delta_\ve(\bs u)=\sqrt{\ve}c_3(\bs u)+o(\sqrt{\ve}),
$
where $c_3$ is odd. 
We compute the moments of the truncated prefactor $J_{\ve,\omega}$ from \cref{sec:mod-ibis-general} and, for brevity, write $J_\ve$ for it in this subsection. For $\bs u\in\mathcal{B}_R$, the event within the cutoff is exactly $\{s\geq s_\ve^\ast(\bs u)\}$; since $|\delta_\ve|=O(\sqrt{\ve})$ on $\mathcal{B}_R$, the cutoff does not affect the part of the event below the parabola and changes the top-branch integrals below only by exponentially small amounts.
Define
\[
(\delta_\ve)_+:=\max\{\delta_\ve,0\},\qquad (\delta_\ve)_-:=\max\{-\delta_\ve,0\},
\]
and
\[
m_1:=\E_R[(c_3(\bs u))_+]=\E_R[(c_3(\bs u))_-].
\]
The equality follows because $c_3$ is odd and both the Gaussian distribution and $\mathcal{B}_R$ are symmetric. We now first derive the leading mean and variance of $J_\ve$ as defined in~\eqref{eq:mod-ibis-def}. Since $h_\ve(s)=1+O(\ve)$ in the region contributing to the leading moments, we omit it in the following leading-order calculation. Set $t=s-b(\bs u)$. Conditional on $\bs u$, the event is $t\geq\delta_\ve(\bs u)$, and hence
\[
\E[J_\ve\mid\bs u]=\int_{\delta_\ve(\bs u)}^\infty e^{-t}\,\mathrm{d}t+o(\sqrt{\ve})=e^{-\delta_\ve(\bs u)}+o(\sqrt{\ve}).
\]
Since $\delta_\ve=\sqrt{\ve}c_3+o(\sqrt{\ve})$ and $\E_R[c_3]=0$, it follows that
\[
\E_R[J_\ve]=1+o(\sqrt{\ve}).
\]
For the second moment, we consider the branches above and below the parabola separately. On the top branch, $t\geq0$, and the event requires $t\geq(\delta_\ve)_+$. Thus,
\[
\E[J_\ve^2\mid\bs u,\text{ top contribution}]=\frac{1}{1-w_\ve}\int_{(\delta_\ve)_+}^\infty e^{-t}\,\mathrm{d}t=\frac{e^{-(\delta_\ve)_+}}{1-w_\ve}.
\]
On the bottom branch, write $v=b(\bs u)-s=-t\geq0$. This branch contributes only when $\delta_\ve<0$, in which case $0\leq v\leq(\delta_\ve)_-$. Therefore,
\[
\E[J_\ve^2\mid\bs u,\text{ bottom contribution}]=\frac{1}{w_\ve}\int_0^{(\delta_\ve)_-}\frac{e^{2v}}{g_\ve(v)}\,\mathrm{d}v.
\]
Define
\[
I_{g_\ve}(t):=\int_0^t\frac{e^{2v}}{g_\ve(v)}\,\mathrm{d}v,\qquad A_\ve(g_\ve):=\E_R[I_{g_\ve}((\delta_\ve)_-)].
\]
Combining the two branches gives
\[
\E_R[J_\ve^2]=\frac{\E_R[e^{-(\delta_\ve)_+}]}{1-w_\ve}+\frac{A_\ve(g_\ve)}{w_\ve}+o(\sqrt{\ve}).
\]
Since
\[
(\delta_\ve)_+=\sqrt{\ve}(c_3(\bs u))_++o(\sqrt{\ve}),
\]
we have
\[
\E_R[e^{-(\delta_\ve)_+}]=1-m_1\sqrt{\ve}+o(\sqrt{\ve}).
\]
Using $(1-w_\ve)^{-1}=1+w_\ve+O(w_\ve^2)$ and $\E_R[J_\ve]=1+o(\sqrt{\ve})$, we obtain the general variance expansion
\[
\Var_R(J_\ve)=w_\ve+\frac{A_\ve(g_\ve)}{w_\ve}-m_1\sqrt{\ve}+O(w_\ve^2+w_\ve\sqrt{\ve})+o(\sqrt{\ve}).
\]

The quantity $A_\ve(g_\ve)$ can also be written using the tail distribution of $(\delta_\ve)_-$. Define
\[
S_\ve(v):=\PP_R((\delta_\ve(\bs u))_->v)=\PP_R(\delta_\ve(\bs u)<-v).
\]
By Fubini's theorem,
\begin{align*}
A_\ve(g_\ve)
&=\E_R\left[\int_0^{(\delta_\ve)_-}\frac{e^{2v}}{g_\ve(v)}\,\mathrm{d}v\right]=\int_0^\infty\frac{e^{2v}}{g_\ve(v)}\PP_R((\delta_\ve)_->v)\,\mathrm{d}v=\int_0^\infty\frac{S_\ve(v)e^{2v}}{g_\ve(v)}\,\mathrm{d}v.
\end{align*}

\begin{proof}[Proof of \cref{thm:scaled-left-proposal}]
For the rescaled proposal density
$
g_\ve(v)=\ve^{-1/2} g\left(\ve^{-1/2} v\right),
$
a change of variables gives
\begin{align*}
I_{g_\ve}(\sqrt{\ve}y)
&=\int_0^{\sqrt{\ve}y}\frac{e^{2v}}{\ve^{-1/2}g(v/\sqrt{\ve})}\,\mathrm{d}v=\ve\int_0^y\frac{e^{2\sqrt{\ve}x}}{g(x)}\,\mathrm{d}x=\ve\int_0^y\frac{1}{g(x)}\,\mathrm{d}x+o(\ve).
\end{align*}
Since $(\delta_\ve(\bs u))_-/\sqrt{\ve}\to(c_3(\bs u))_-$ uniformly on $\mathcal{B}_R$, we obtain
\begin{align*}
A_\ve(g_\ve)
&=\E_R[I_{g_\ve}((\delta_\ve(\bs u))_-)]=\ve\E_R\left[\int_0^{(c_3(\bs u))_-}\frac{1}{g(x)}\,\mathrm{d}x\right]+o(\ve)=D_g\ve+o(\ve).
\end{align*}
For $w_\ve=\alpha\sqrt{\ve}$,
\[
\frac{\E_R[e^{-(\delta_\ve)_+}]}{1-w_\ve}=\left(1-m_1\sqrt{\ve}+o(\sqrt{\ve})\right)\left(1+\alpha\sqrt{\ve}+O(\ve)\right)=1+(\alpha-m_1)\sqrt{\ve}+o(\sqrt{\ve}),
\]
and
\[
\frac{A_\ve(g_\ve)}{w_\ve}=\frac{D_g\ve+o(\ve)}{\alpha\sqrt{\ve}}=\frac{D_g}{\alpha}\sqrt{\ve}+o(\sqrt{\ve}).
\]
Since $\E_R[J_\ve]^2=1+o(\sqrt{\ve})$, it follows that
\[
\Var_R(J_\ve)=\left(\alpha+\frac{D_g}{\alpha}-m_1\right)\sqrt{\ve}+o(\sqrt{\ve}).
\]
The leading constant is positive: if $c_3\equiv0$, then $m_1=D_g=0$ and it equals $\alpha>0$. Otherwise, $D_g>0$, and by the Cauchy--Schwarz inequality,
\begin{align*}
m_1&=\E_R\left[\int_0^{(c_3(\bs u))_-}\sqrt{g(x)}\,\frac{1}{\sqrt{g(x)}}\,\mathrm{d}x\right]\leq\left(\E_R\left[\int_0^{(c_3(\bs u))_-}g(x)\,\mathrm{d}x\right]\right)^{1/2}D_g^{1/2}\\
&\leq\left(\PP_R(c_3(\bs u)<0)\,D_g\right)^{1/2}\leq\sqrt{D_g/2},
\end{align*}
since $\int_0^\infty g=1$ and $\PP_R(c_3(\bs u)<0)=\PP_R(c_3(\bs u)>0)\leq\frac12$ by symmetry. Hence $\alpha+D_g/\alpha-m_1\geq(2-2^{-1/2})\sqrt{D_g}>0$. Therefore,
\[
\Cov_R(J_\ve)=\frac{\sqrt{\Var_R(J_\ve)}}{\E_R[J_\ve]}=\left(\alpha+\frac{D_g}{\alpha}-m_1\right)^{1/2}\ve^{1/4}+o(\ve^{1/4}).
\]
If $D_g > 0$, the terms depending on $\alpha$ satisfy
\[
\alpha+\frac{D_g}{\alpha}\geq2\sqrt{D_g},
\]
with equality at $\alpha = \sqrt{D_g}$. Thus, the asymptotically optimal mixture weight within this scaled family is 
\[
w_\ve^\ast\sim\sqrt{D_g\ve},\qquad \Var_R(J_\ve)\sim(2\sqrt{D_g}-m_1)\sqrt{\ve}.
\]
If $D_g = 0$, then $c_3\equiv 0$ and $m_1 = 0$, so the variance expansion instead reduces to
\[\Var_R(J_\ve) = \alpha \sqrt{\ve} + o(\sqrt{\ve})\]
for every fixed $\alpha > 0$. There is no minimizing positive $\alpha$ at this order; determining an optimal scaling requires higher-order information about the boundary discrepancy.
\end{proof}

\noindent\textit{On the restriction to $\mathcal{B}_R$.} We justify the statements following~\eqref{eq:scaled-left-density}. Throughout this discussion, expectations are taken under the full Gaussian law of $\bs u$, while the normal cutoff in $J_{\ve, \omega}$ is retained.

First, let $g(x)=\beta e^{-\beta x}$, $w_\ve=\alpha\sqrt{\ve}$, and suppose $c_3\not\equiv0$. Since $c_3$ is an odd homogeneous cubic polynomial, there are $c_0>0$ and a set $S$ of unit vectors of positive surface measure with $c_3(\hat{\bs u})\leq-c_0$ for $\hat{\bs u}\in S$. For $\bs u=\rho\hat{\bs u}$ with $\hat{\bs u}\in S$ and $\rho\in[\ve^{-1/4},\ve^{-1/4}+1]$, \cref{lem:boundary-expansion} gives $(\delta_\ve(\bs u))_-\geq\sqrt{\ve}\,c_0\rho^3-C\ve(1+\rho)^4\geq c_0\ve^{-1/4}-C'$, and the corresponding bottom-branch points satisfy $|\ve s|=O(\sqrt{\ve})$, so that they lie within the cutoff and $h_\ve(s)$ is bounded below. 

Since $1/g_\ve(v)=\sqrt{\ve}\,e^{\beta v/\sqrt{\ve}}/\beta$, the bottom-branch contribution to $\E[J_{\ve,\omega}^2\mid\bs u]$ is at least $c'\sqrt{\ve}\,\big(e^{\beta(\delta_\ve(\bs u))_-/\sqrt{\ve}}-1\big)$. The Gaussian probability of this set of $\bs u$ is at least $c'e^{-(\ve^{-1/4}+1)^2/2}$, and hence $\Var(J_{\ve,\omega})\geq\exp\big(\beta c_0\ve^{-3/4}-C''\ve^{-1/2}\big)\to\infty$ as $\ve\downarrow0$; the half-normal case is analogous.

Second, write $\bs\eta_\ve(s,\bs u)-\bs\eta_z=d_\parallel\bs e_z+\bs d_\perp$, i.e., $d_\parallel=\ve s/r$ and $\bs d_\perp=\sqrt{\ve}\,G_z^{-1/2}\bs u$. The bottom-branch part of $\E[J_\ve^2]$ is
\[
\int_{\bs\eta_z^\perp}\int_{-\infty}^{b(\bs u)}\varphi_\perp(\bs u)\,\frac{e^{2v}h_\ve(s)^2}{w_\ve g_\ve(v)}\,\ind_{\{F(\bs\eta_\ve(s,\bs u))\geq z\}}\,\mathrm{d}s\,\mathrm{d}\bs u,\qquad v=b(\bs u)-s,
\]
and inserting $b(\bs u)=-\frac{\lambda_z}{2\ve}\langle\bs d_\perp,H_z\bs d_\perp\rangle$ and $\Vert\bs u\Vert^2=\frac1\ve\langle\bs d_\perp,G_z\bs d_\perp\rangle$, a direct computation gives
\[
-\frac12\Vert\bs u\Vert^2+2v-\frac{\ve s^2}{r^2}=-\frac{1}{\ve}Q(\bs\eta),\qquad Q(\bs\eta):=\Vert\bs\eta\Vert^2-r^2-\frac12\langle\bs d_\perp,G_z\bs d_\perp\rangle.
\]
Thus, up to the factor $1/(w_\ve g_\ve(v))$, event points below the parabola contribute $e^{-Q(\bs\eta)/\ve}$. Near $\bs\eta_z$, $Q(\bs\eta)\approx\frac12\langle\bs d_\perp,G_z\bs d_\perp\rangle>0$ on the event, but \cref{assump:standard-ldt} does not ensure that $Q$ is bounded away from zero on the event outside a neighborhood of $\bs\eta_z$; for $R=\infty$, this has to be assumed in addition to a heavy-tailed $g$. In contrast, the same computation for the top branch gives the exponent $-\frac1\ve\big(\frac12\Vert\bs d_\perp\Vert^2+rd_\parallel+d_\parallel^2\big)\leq-\frac{1}{2\ve}\langle\bs d_\perp,G_z\bs d_\perp\rangle$ for $s\geq b(\bs u)$, without further assumptions.

\begin{proof}[Derivation of \cref{rem:unscaled-left-proposal}]
Suppose $g_\ve=g$ does not depend on $\ve$, where $g$ is continuous and positive at zero. Since
$
e^{2v} / g(v)\longrightarrow 1 / g(0)
$ as $v\downarrow0$,
we have
$
I_g(t)=t / g(0) +o(t).
$
Therefore,
\begin{align*}
A_\ve(g)
&=\frac{\E_R[(\delta_\ve(\bs u))_-]}{g(0)}+o(\sqrt{\ve})=\frac{m_1}{g(0)}\sqrt{\ve}+o(\sqrt{\ve}).
\end{align*}
Taking $w_\ve=\alpha\ve^{1/4}$ in the general variance expansion gives
\[
\Var_R(J_\ve)=\left(\alpha+\frac{m_1}{\alpha g(0)}\right)\ve^{1/4}+o(\ve^{1/4}).
\]
Consequently,
\[
\Cov_R(J_\ve)=\left(\alpha+\frac{m_1}{\alpha g(0)}\right)^{1/2}\ve^{1/8}+o(\ve^{1/8}).
\]
If $m_1 > 0$, the leading constant is minimized at 
$
\alpha=\sqrt{m_1 / g(0)},
$
so
\[
w_\ve^\ast\sim\sqrt{\frac{m_1}{g(0)}}\,\ve^{1/4},\qquad \Var_R(J_\ve)\sim2\sqrt{\frac{m_1}{g(0)}}\,\ve^{1/4}.
\]
If $m_1 = 0$, the variance expansion remains valid for every fixed $\alpha > 0$, with leading coefficient $\alpha$, but the optimal scaling is not determined at this order.
\end{proof}

We finally derive the formally optimal bottom-branch density stated in \cref{rem:optimal-left-distribution}. For fixed $\ve$, the terms in the leading variance that depend on $w_\ve$ and $g_\ve$ are
\[
\Phi_\ve(w_\ve,g_\ve):=w_\ve+\frac{A_\ve(g_\ve)}{w_\ve},\qquad A_\ve(g_\ve)=\int_0^\infty\frac{S_\ve(v)e^{2v}}{g_\ve(v)}\,\mathrm{d}v.
\]
For fixed $g_\ve$, the arithmetic-geometric mean inequality gives
\[
\Phi_\ve(w_\ve,g_\ve)\geq2\sqrt{A_\ve(g_\ve)},
\]
with equality at $w_\ve=\sqrt{A_\ve(g_\ve)}$, provided $0 < A_\ve(g_\ve) < 1$. It remains to minimize $A_\ve(g_\ve)$ over all probability densities $g_\ve$. Since $\int_0^\infty g_\ve(v)\,\mathrm{d}v=1$, the Cauchy--Schwarz inequality gives
\begin{align*}
A_\ve(g_\ve)
&=\left(\int_0^\infty\frac{S_\ve(v)e^{2v}}{g_\ve(v)}\,\mathrm{d}v\right)\left(\int_0^\infty g_\ve(v)\,\mathrm{d}v\right)\geq\left(\int_0^\infty e^v\sqrt{S_\ve(v)}\,\mathrm{d}v\right)^2.
\end{align*}
Define
\[
\mathcal{Z}_\ve:=\int_0^\infty e^v\sqrt{S_\ve(v)}\,\mathrm{d}v.
\]
Equality holds when
$
g_\ve^\ast(v)=e^v\sqrt{S_\ve(v)} / \mathcal{Z}_\ve.
$
If $0 < \mathcal{Z}_\ve < 1$, this density gives $A_\ve(g_\ve^\ast)=\mathcal{Z}_\ve^2$, and the corresponding optimal weight for the leading-order objective is
$
w_\ve^\ast=\sqrt{A_\ve(g_\ve^\ast)}=\mathcal{Z}_\ve.
$
On the fixed region $\mathcal{B}_R$, $\mathcal{Z}_\ve = O(\sqrt{\ve})$, so the upper constraint is satisfied for sufficiently small $\ve$. If $\mathcal{Z}_\ve = 0$, there is no bottom-branch event contribution, and the infimum of this objective is approached as $w_\ve \downarrow 0$.

\section{Derivations and proofs for KL-divergence analysis}\label{app:proofs-sec4}
Here, we summarize the derivations and proofs for \cref{sec:theory-kl}. 

\subsection{Proof sketch for \cref{thm:chatter-diac}} We sketch the arguments from \cite{chatterjee-diaconis:2018} for the simplified formulation of the theorem.
The main idea is to decouple the concentration events: Define the benign event $G_a(\bs\eta)=\{\rho_{\ve,z}(\bs\eta)\leq a\}$ for some $a>0$, and the corresponding malign counterpart $G^c_a(\bs\eta)=\{\rho_{\ve,z}(\bs\eta)> a\}$ with the auxiliary function $h(\bs\eta)=\mathbbm{1}_{G_a(\bs\eta)}$. Define ${\cal I}_M(f)\coloneqq\frac{1}{M}\sum_{i=1}^M\rho_{\ve,z}(\bs\eta^{(i)})f(\bs\eta^{(i)})$ and ${\cal I}(f)\coloneqq\mathbb{E}[{\cal I}_M(f)]=\mathbb{E}_{\pi_{\ve,z}}[f]$ for any $f$. To prove (i), we observe
\begin{equation*}
    \frac{\left|\hat{P}_{M,\ve}(z)-P_\ve(z)\right|}{\mathbb{E}[\hat{P}_{M,\ve}(z)]}=\left|{\cal I}_M(1)-{\cal I}(1)\right|\leq\underbrace{\left|{\cal I}_M(1)-{\cal I}_M(h)\right|}_{\eqqcolon(a)}+\underbrace{\left|{\cal I}_M(h)-{\cal I}(h)\right|}_{\eqqcolon(b)}+\underbrace{\left|{\cal I}(h)-{\cal I}(1)\right|}_{\eqqcolon(c)}\,.
\end{equation*}
Terms $(a)$ and $(c)$ can be upper bounded in expectation via
\begin{equation*}
    \mathbb{E}\left[\left|{\cal I}_M(1)-{\cal I}_M(h)\right|\right]\leq0\cdot\mathbb{E}_{\pi_{\ve,z}}\left[\mathbbm{1}_{G_a(\bs\eta)}\right]+1\cdot\mathbb{E}_{\pi_{\ve,z}}\left[\mathbbm{1}_{G^c_a(\bs\eta)}\right]=\mathbb{P}_{\pi_{\ve,z}}\left[G^c_a(\bs\eta)\right]\,.
\end{equation*}
Term $(b)$ controls the stochastic error via
\begin{equation*}
    \mathbb{E}\left[\left|{\cal I}_M(h)-{\cal I}(h)\right|\right]^2\leq\Var\left({\cal I}_M(h)\right)\leq\frac{a}{M}\mathbb{P}_{\pi_{\ve,z}}\left[G_a(\bs\eta)\right]\leq\frac{a}{M}\,.
\end{equation*}
Choosing $a=\exp\{\mathrm{KL}(\pi_{\ve,z}\|\mu_\ve)+t\}$ and $M=\exp\{\mathrm{KL}(\pi_{\ve,z}\|\mu_\ve)+2t\}$ for $t>0$ recovers~\eqref{eq:cd-suff}.
To prove (ii), observe for $\zeta>0$, since $\hat{P}_{M,\ve}(z)/P_\ve(z)={\cal I}_M(1)$, and via Markov's inequality
\begin{align*}
    \mathbb{P}\left[{\cal I}_M(1)\geq\zeta\right]
    \leq\sum_{i=1}^M\mathbb{P}_{\mu_\ve}\left[G^c_a(\bs\eta^{(i)})\right]+\mathbb{P}\left[{\cal I}_M(h)\geq\zeta\right]\leq\frac{M}{a}+\frac{1}{\zeta}\mathbb{P}_{\pi_{\ve,z}}\left[G_a(\bs\eta)\right]\,.
\end{align*}
Choosing $a=\exp\{\mathrm{KL}(\pi_{\ve,z}\|\mu_\ve)-t\}$ and $M=\exp\{\mathrm{KL}(\pi_{\ve,z}\|\mu_\ve)-2t\}$ for $t>0$ recovers~\eqref{eq:cd-nece}.

\subsection{Proof of \cref{prop:vanilla-ibis-KL}: Vanilla IBIS KL divergence}

For $\bs \eta$ with $F(\bs \eta) \geq z$, we have
$\log\rho_{\ve,z}(\bs\eta)=-\log P_\ve(z)-I(z)/\ve-rs$
and hence
\begin{equation*}
    \mathrm{KL}(\pi_{\ve,z}\|\mu_\ve)=\mathbb{E}_{\pi_{\ve,z}}\left[\log\rho_{\ve,z}(\bs\eta)\right]=-\log P_\ve(z)-I(z)/\ve-r\mathbb{E}_{\pi_{\ve,z}}\left[s\right]\,.
\end{equation*}
The first two terms on the RHS have the limiting behavior
\begin{align*}
    -\log P_\ve(z)-I(z)/\ve = \frac{1}{2} \log(1/\ve) + \frac{1}{2} \log(2 \pi) - \log C_{\rm SORM}(z, \lambda_z) + o(1)\,,
\end{align*}
and it only remains to compute $-r\mathbb{E}_{\pi_{\ve,z}}\left[s\right]$ as $\ve \downarrow 0$. To do this, we write
\begin{align*}
    \pi_{\ve}(\bs\eta)\mathrm{d}\bs\eta&=(2\pi\ve)^{-N/2}e^{-\|\bs\eta\|^2/2\ve}\mathrm{d}\bs\eta
    =(2\pi)^{-\frac{N}{2}}\ve^{\frac{1}{2}}e^{-\frac{I(z)}{\ve}}\exp\left\{-rs-\frac{\ve s^2}{2}-\frac{\|\bs u\|^2}{2}\right\}\mathrm{d}s\mathrm{d}\bs u\,,
\end{align*}
and hence the law conditioned on the rare event in these coordinates becomes
\begin{align*}
    \pi_{\ve,z}(\bs\eta)\mathrm{d}\bs\eta&=\frac{\ind_{\{F(\bs \eta) \geq z\}}}{P_\ve(z)}\pi_{\ve}(\bs\eta)\mathrm{d}\bs\eta
    = (2\pi)^{-\frac{N-1}{2}}\; C^{-1}_{\rm SORM}(z, \lambda_z) \; e^{-rs-\frac{\|\bs u\|^2}{2}}\ind_{ \{s \geq \tau( {\bs u})\}}\mathrm{d}s\mathrm{d}\bs u+o(1)\,.
\end{align*}
The expectation is then given by
\begin{align*}
    \mathbb{E}_{\pi_{\ve,z}}\left[s\right]
    &=(2\pi)^{-\frac{N-1}{2}}\; C^{-1}_{\rm SORM}(z, \lambda_z) \; \int_{\bs\eta_z^\perp}\;\left[\int_{\mathbb{R}} s e^{-rs }\ind_{\{s\geq\tau(\bs u)\}}\mathrm{d}s\right]e^{-\frac{\|\bs u\|^2}{2}}\mathrm{d}\bs u+o(1)\,.
\end{align*}
Calculating first the exponential integral over $s$, and then the Gaussian integrals over $\bs u$, directly leads to the expression for the KL divergence in \cref{prop:vanilla-ibis-KL}.

For the variance of the log weight, note that $\Var_{\pi_{\ve,z}}(\log\rho_{\ve,z})=\Var_{\pi_{\ve,z}}(rs)$. Under the limiting conditional law displayed above, the marginal density of $\bs u$ is proportional to $e^{-\Vert\bs u\Vert^2/2-r\tau(\bs u)}=e^{-\langle\bs u,G_z\bs u\rangle/2}$, i.e., $\bs u\sim\mathcal{N}(\bs 0,G_z^{-1})$, and, given $\bs u$, $E:=r(s-\tau(\bs u))\sim\mathrm{Exp}(1)$. Hence $rs=r\tau(\bs u)+E$ with $E$ independent of $\bs u$, and
\begin{equation*}
\Var(rs)=\Var(E)+\tfrac14\Var\left(\langle\bs u,\lambda_zH_z\bs u\rangle\right)=1+\tfrac12\operatorname{tr}\left(\left(\lambda_zH_zG_z^{-1}\right)^2\right).
\end{equation*}
Convergence of the first and second moments of $s$ under $\pi_{\ve,z}$ to those of the limiting law follows analogously to the proof of \cref{prop:vanilla-ibis-moments}.

\subsection{Proof of \cref{lem:vanilla-ibis-comp}}
Recall $M_z\coloneqq\lambda_zH_zG^{-1}_z$. Then,
\begin{align*}
    \operatorname{Id}_{\bs \eta_z^\perp}-M_z &= (G_z-\lambda_zH_z)G^{-1}_z=(\operatorname{Id}_{\bs \eta_z^\perp} - 2\lambda_z H_z )G^{-1}_z, \quad \text{and thus} \\ 
    \det\left(\operatorname{Id}_{\bs \eta_z^\perp}-M_z\right) &={\det \left( \operatorname{Id}_{\bs \eta_z^\perp} - 2\lambda_z H_z \right)}/{\det(G_z)},
\end{align*}
so that $C_{\rm SORM}(z,\lambda_z)/C_{\rm SORM}(z,2\lambda_z)=\det(\operatorname{Id}_{\bs \eta_z^\perp}-M_z)^{1/2}$. Therefore,
\begin{equation*}
     \frac{\beta_{\rm KL}}{\beta_{\rm CoV}}=\frac{2}{e}\,e^{\frac12\tr(M_z)}\,\frac{C_{\rm SORM}(z,\lambda_z)}{C_{\rm SORM}(z,2\lambda_z)}=\frac{2}{e}\left(\det\left(\operatorname{Id}_{\bs \eta_z^\perp}-M_z\right)e^{\tr(M_z)}\right)^{\frac{1}{2}}=\frac{2}{e}\left(\prod_{i=1}^{N-1}(1-\xi_i)e^{\xi_i}\right)^{\frac{1}{2}}
\end{equation*}
with $\xi_i\coloneqq\frac{\kappa_i}{1-\kappa_i}$ the eigenvalues of $M_z$. By \cref{assump:variance}(ii), $\kappa_i<\frac{1}{2}$ and hence $\xi_i<1$; since $(1-\xi)e^{\xi}\leq 1$ for all $\xi<1$ (with equality only at $\xi=0$), the bound $\frac{\beta_{\rm KL}}{\beta_{\rm CoV}}\leq\frac{2}{e}$ follows.

\subsection{Proof of \cref{prop:mod-ibis-KL}}

We write $J_\ve(s,\bs u)=\ind_{\{F(\bs \eta_\ve(s,\bs u))\geq z\}}h_\ve(s)\tilde{h}_\ve(s,\bs u)$ with
\begin{equation*}
    \tilde{h}_\ve(s,\bs u)\coloneqq\frac{1}{1-w_\ve}\ind_{\{s\geq b(\bs u)\}}+\frac{1}{w_\ve}\frac{\exp\{-(s-b(\bs u))\}}{g_\ve(b(\bs u)-s)}\ind_{\{s < b(\bs u)\}}\,.
\end{equation*}
A short computation then yields $\log\rho_{\ve,z}=\log(D_\ve/P_\ve(z))+\log h_\ve(s)+\log\tilde{h}_\ve(s,\bs u)$ on the event set, with $D_\ve = \sqrt{\ve / (2 \pi)} C_{\rm SORM} (z, \lambda_z) e^{-I(z) / \ve}$ as in \cref{tab:ibis-family}. Expanding in $\ve$, with analogous arguments as before, we find $\operatorname{KL}(\pi_{\ve,z} \| \mu_\ve) = \E_{\pi_{\ve,z}}[\log\tilde{h}_\ve] + O(\ve)$.

Let $\Omega_\ve:=\{|\ve s|\leq\delta_0,\ \Vert\bs u\Vert\leq\delta_0/\sqrt{\ve}\}$, with $\delta_0$ from \cref{lem:boundary-expansion} in curvature-adjusted coordinates. Outside $\Omega_\ve$, we use $\log(w_\ve g_\ve(v))=\log\alpha+\log g(v/\sqrt{\ve})$ and the growth condition on $g$, which show that $|\log\tilde{h}_\ve(s,\bs u)|$ is bounded by a polynomial in $|s|$, $\Vert\bs u\Vert$, and $\ve^{-1/2}$. Since $\Omega_\ve^c$ lies outside a fixed neighborhood of $\bs\eta_z$, \cref{lemma:uniform-separation} with $\bs p=\bs 0$, together with the Gaussian tails of $\pi_\ve$, shows that $\Omega_\ve^c$ contributes only $O(e^{-c/\ve})$ to $\E_{\pi_{\ve,z}}[\log\tilde{h}_\ve]$ and is hence negligible.

Inside $\Omega_\ve$, the event is $\{s\geq s_\ve^\ast(\bs u)\}$ by \cref{lem:boundary-expansion}, and the cutoff $|\ve s|\leq\delta_0$ only removes an exponentially small part of the top branch. We decompose into the top branch, $t=s-b(\bs u)\geq(\delta_\ve)_+$, and the bottom branch, $0\leq v=b(\bs u)-s\leq(\delta_\ve)_-$, and substitute $v=\sqrt{\ve}x$ for the latter. Up to the factors $(1 + O(\ve))$, the inner $s$-integral of $\log\tilde{h}_\ve$ then equals
\begin{equation*}
\Phi_\ve(\bs u):=-\log(1-w_\ve)\,e^{-(\delta_\ve(\bs u))_+}+\sqrt{\ve}\int_0^{y_\ve(\bs u)}\left(\sqrt{\ve}x-\log\alpha+\log\frac{1}{g(x)}\right)e^{\sqrt{\ve}x}\,\mathrm{d}x,
\end{equation*}
with $y_\ve(\bs u):=(\delta_\ve(\bs u))_-/\sqrt{\ve}$.
For $\Vert\bs u\Vert\leq\ve^{-1/8}$, the first bound in \cref{lem:boundary-expansion} gives $y_\ve(\bs u)\to(c_3(\bs u))_-$ and $y_\ve(\bs u)\leq C(1+\Vert\bs u\Vert)^4$, so that $\sqrt{\ve}\,y_\ve(\bs u)$ is $O(\sqrt{\ve}(1+\Vert\bs u\Vert)^4)$, hence bounded and $O(\sqrt{\ve})$ for every fixed $\bs u$. By the growth condition on $g$, $|\Phi_\ve(\bs u)|\leq C\sqrt{\ve}(1+\Vert\bs u\Vert)^{4(p+1)}$, and dominated convergence gives
\begin{equation*}
\frac{1}{\sqrt{\ve}}\,\E\big[\Phi_\ve(\bs u)\ind_{\{\Vert\bs u\Vert\leq\ve^{-1/8}\}}\big]\longrightarrow\alpha+\E\left[\int_0^{(c_3(\bs u))_-}\left(\log\frac{1}{g(x)}-\log\alpha\right)\mathrm{d}x\right]=\alpha+\widetilde{D}_g-m_1\log\alpha.
\end{equation*}
For larger $\bs u$ with $\ve^{-1/8}<\Vert\bs u\Vert\leq\delta_0/\sqrt{\ve}$, the second bound in \cref{lem:boundary-expansion} gives $\sqrt{\ve}\,y_\ve(\bs u)\leq C(\sqrt{\ve}+\delta_0)(1+\Vert\bs u\Vert)^2$, so that $|\Phi_\ve(\bs u)|$ is bounded by a polynomial in $\ve^{-1/2}$ and $\Vert\bs u\Vert$ times $e^{C(\sqrt{\ve}+\delta_0)(1+\Vert\bs u\Vert)^2}$. Collecting terms yields the first claim of \cref{prop:mod-ibis-KL}:
\begin{equation*}
    \mathrm{KL}(\pi_{\ve,z}\|\mu_\ve)=\sqrt{\ve}\left(\alpha+\widetilde{D}_g-m_1\log\alpha\right)+o(\sqrt{\ve})\,.
\end{equation*}
For the variance of the log weight, we repeat this computation for $\E_{\pi_{\ve,z}}[(\log\rho_{\ve,z})^2]$. The contributions of $\log(D_\ve/P_\ve(z))$ and $\log h_\ve$ are $o(\sqrt{\ve})$ by the above and the Cauchy--Schwarz inequality. On the top branch, $(\log\tilde{h}_\ve)^2=\log(1-w_\ve)^2=O(\ve)$. On the bottom branch, the inner integral becomes $\sqrt{\ve}\int_0^{y_\ve(\bs u)}(\sqrt{\ve}x-\log\alpha+\log\frac{1}{g(x)})^2e^{\sqrt{\ve}x}\,\mathrm{d}x$, which is bounded by $C\sqrt{\ve}(1+\Vert\bs u\Vert)^{4(2p+1)}$ for $\Vert\bs u\Vert\leq\ve^{-1/8}$, while the remaining regions are negligible as before. Dominated convergence gives $\E_{\pi_{\ve,z}}[(\log\rho_{\ve,z})^2]=\sqrt{\ve}\,\E[\int_0^{(c_3(\bs u))_-}(\log\frac{1}{g(x)}-\log\alpha)^2\mathrm{d}x]+o(\sqrt{\ve})$, and since $\mathrm{KL}(\pi_{\ve,z}\|\mu_\ve)^2=O(\ve)$, the second claim follows.

\section{Optimal bottom-branch sampling in a two-dimensional example}
\label{app:toy-example}
This example illustrates the optimal bottom-branch density derived in \cref{rem:optimal-left-distribution}. The event boundary is explicit, allowing us to evaluate this density and sample from it numerically. We use it as a benchmark to examine the shape of the optimal density and compare its sampling performance with simpler half-normal and exponential choices. Here, optimality refers to the leading-order variance criterion within the proposed sampling family.
Let 
\begin{equation*}
F_\kappa (x_1, x_2) = x_2 - \cosh x_1 + \kappa x_1^3, \quad z = \kappa = 0.1,
\end{equation*}
and estimate $P_\ve(z) = \PP(F_\kappa(\sqrt{\ve}\bs \xi) \geq z)$, where $\bs \xi \sim \mathcal{N}(\bs 0, \Id_2)$. The instanton is $\bs \eta_z = (0, 1.1)$, with $r = \lambda_z = 1.1$, $I(z) = 0.605$, and $\nabla^2 F_\kappa(\bs \eta_z) = \text{diag}(-1, 0)$. Thus, $G_z = 2.1$. The cubic term allows the exact event boundary to extend below its quadratic approximation, making the branch below the parabola necessary.

\begin{figure}
    \centering
    \includegraphics{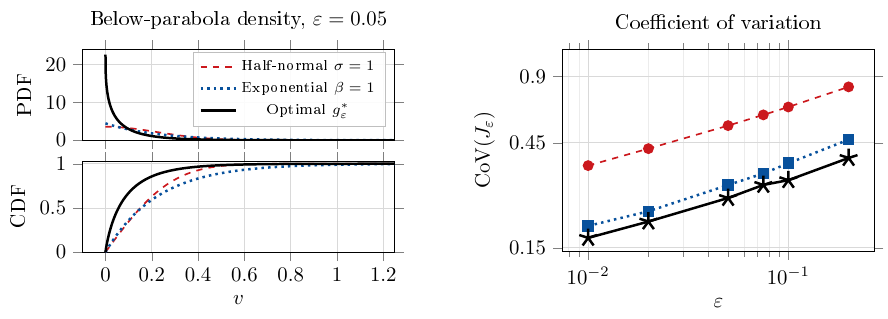}
    \vspace{-2ex}
\caption{Comparison of different proposals $g$ for the two-dimensional toy problem. Left: CDFs of the scaled half-normal proposal with $\sigma = 1$, the scaled exponential proposal with $\beta = 1$, and the optimal proposal $g_\ve^\ast$, at $\ve = 0.05$. Right: the corresponding coefficients of variation over $\ve \in [0.01, 0.20]$, estimated using $M = 10^6$ samples.}
\label{fig:toy-cdf-cov}
\end{figure}

In the coordinates $x_1 = \sqrt{\ve/G_z}u$, $x_2 = r + \ve s/r$, with $u \sim \mathcal{N}(0,1)$, the event is $s \geq s_\ve^\ast(u) = b(u) + \delta_\ve(u)$, where
\begin{equation*}
b(u) = \frac{r}{2G_z}u^2, \quad \delta_\ve(u) = \frac{r}{\ve}\left[ \cosh\left( \sqrt{\frac{\ve}{G_z}}u \right) -1 - \frac{\ve u^2}{2 G_z}\right] - \frac{\kappa r\sqrt{\ve}}{G_z^{3/2}}u^3.
\end{equation*}
Set $v_{\max} = -{\min}_u \delta_\ve(u)$. For $0 < v < v_{\max}$, the equation $\delta_\ve(u) = -v$ has two positive roots $u_-(v) < u_+(v)$. Consequently,
\begin{equation*}
S_\ve(v) = \Phi(u_+(v)) - \Phi(u_-(v)),
\end{equation*}
where $\Phi$ is the standard normal CDF; $S_\ve(v) = 0$ for $v \geq v_{\max}$. The proposal in \cref{rem:optimal-left-distribution} therefore becomes
\begin{equation*}
\mathcal{Z}_\ve = \int_0^{v_{\max}} e^v \sqrt{S_\ve(v)}\,dv, \quad g_\ve^\ast(v) = \frac{e^v \sqrt{S_\ve(v)}}{\mathcal{Z}_\ve}, \quad w_\ve^\ast = \cal Z_\ve.
\end{equation*}
Evaluating this density requires only scalar root finding and one-dimensional quadrature. We sample from it by numerically inverting its CDF $H_\ve(v) = \int_0^v g_\ve^\ast(t)\,dt$.

We use $M = 10^6$ samples for $\ve \in [0.01, 0.20]$ and compare $g_\ve^\ast$ with a half-normal proposal with $\sigma = 1$ and an exponential proposal with $\beta = 1$, both scaled by $\sqrt{\ve}$. The CDF comparison in \cref{fig:toy-cdf-cov} shows that $g_\ve^\ast$ places more mass near the quadratic boundary. It gives the lowest measured CoV through the tested range, while the exponential proposal performs nearly as well. Thus, in this example, a simple exponential density achieves nearly the same sampling performance without requiring the computation of $S_\ve$ or numerical inversion of the optimal CDF.

\end{document}